\documentclass[a4paper,11pt]{article}
\usepackage[utf8]{inputenc}
\usepackage[T1]{fontenc}
\usepackage[english]{babel}
\usepackage[hmargin={32mm,32mm},vmargin={32mm,35mm}]{geometry}
\usepackage{amsmath}
\usepackage{amsfonts}
\usepackage{amssymb}
\usepackage{enumerate}
\usepackage{graphicx}
\usepackage{bbm}
\usepackage[dvipsnames]{xcolor}
\usepackage{xspace}
\usepackage{url}
\usepackage[hidelinks]{hyperref}
\usepackage{cite}
\usepackage{listings}
\usepackage{caption}
\usepackage{mathrsfs}
\usepackage{amstext}
\usepackage{array}
\newcolumntype{C}{>{$}c<{$}}

\usepackage{bookmark}

\newcommand{\bra}[1]{\langle #1 |}
\newcommand{\ket}[1]{|#1\rangle}

\newcommand{\eg}{e.g.\@\xspace}
\newcommand{\ie}{i.e.\@\xspace}
\newcommand{\Eq}[1]{Eq.\@\xspace\eqref{#1}}
\newcommand{\Eqs}[1]{Eqs.\@\xspace\eqref{#1}}
\newcommand{\Fig}[1]{Fig.\@\xspace\ref{#1}}

\newcommand{\updown}[2]{^{#1}_{\phantom{#1}#2}}
\newcommand{\downup}[2]{_{#1}^{\phantom{#1}#2}}

\newcommand{\Id}{\mathbbm{1}}
\newcommand{\R}{\mathbbm{R}}

\newcommand{\Rt}{\,{}^{(3)}\!R}
\newcommand{\Et}{\widetilde E}
\newcommand{\EE}{{\cal E}}
\newcommand{\EEt}{\widetilde{\cal E}}

\newcommand{\Gc}{{\Gamma_{\! 0}}}
\newcommand{\Oj}{{\cal O}\bigl(\sqrt j\bigr)}

\newcommand{\D}{{\cal D}}

\newcommand{\DD}{{\mathscr{D}}}
\newcommand{\threedots}{{\lower.1em\hbox{\vdots}}}

\numberwithin{equation}{section}

\begin{document}

\begin{center}

\Large
\textbf{Scalar curvature operator for quantum-reduced loop gravity}

\vspace{16pt}

\large
Jerzy Lewandowski$^1$ and Ilkka Mäkinen$^{2,\; 1}$

\normalsize

\vspace{12pt}

$^1$Faculty of Physics, University of Warsaw \\
Pasteura 5, 02-093 Warsaw, Poland

\vspace{8pt}

$^2$National Centre for Nuclear Research \\
Pasteura 7, 02-093 Warsaw, Poland

\vspace{8pt}

jerzy.lewandowski@fuw.edu.pl, ilkka.makinen@ncbj.gov.pl

\end{center}

\renewcommand{\abstractname}{\vspace{-\baselineskip}}

\begin{abstract}

	\noindent In a previous article we have introduced an operator representing the three-dimensio-nal scalar curvature in loop quantum gravity. In this article we examine the new curvature operator in the setting of quantum-reduced loop gravity. We derive the explicit form of the curvature operator as an operator on the Hilbert space of the quantum-reduced model. As a simple practical example, we study the expectation values of the operator with respect to basis states of the reduced Hilbert space.

\end{abstract}

\section{Introduction}

The three-dimensional Ricci scalar is a geometrical quantity of fundamental importance in the $3+1$ formulation of general relativity, describing the curvature of the spatial surfaces. In addition to its role as a basic geometrical observable, the relevance of the scalar curvature to loop quantum gravity arises also from the fact that the Ricci scalar can enter the definition of the dynamics of the theory through the Lorentzian part of the Hamiltonian constraint operator. Following the well-known construction due to Thiemann \cite{QSD}, the quantization of the Lorentzian part of the Hamiltonian in loop quantum gravity is usually based on its expression in terms of the extrinsic curvature. However, it is also possible to replace this relatively complicated form of the Lorentzian term with a seemingly simpler expression, which is essentially the three-dimensional Ricci scalar integrated over the spatial manifold (see \eg \cite{LS14, paper2}).

In a previous article \cite{part1}, we have introduced a new operator representing the scalar curvature in loop quantum gravity. The operator is constructed within the kinematical framework of loop quantum gravity, but its definition is limited to the Hilbert space of states based on a fixed cubical graph. The classical starting point of the construction is to express the Ricci scalar as a function of the densitized triad and its gauge covariant derivatives. In passing from the classical expression to a well-defined quantum operator, the restriction to a cubical graph plays an essential role, allowing one to regularize the covariant derivatives of the triad in a relatively straightforward manner, in terms of finite differences of parallel transported flux variables associated to neighboring nodes of the graph.

While more work is needed to extend the definition of the new curvature operator to the entire kinematical Hilbert space of loop quantum gravity, which contains states based on all possible graphs, the operator in its present form can be applied to models which are formulated in the kinematical setting of full loop quantum gravity using states defined on cubical graphs. From the point of view of the potential physical applications of the model (see \cite{qrlg-phys-1, qrlg-phys-2, qrlg-phys-3, qrlg-phys-4} for a selection of examples), a particularly interesting approach of this kind is quantum-reduced loop gravity \cite{qrlg1, qrlg2, qrlg3, qrlg4}. Quantum-reduced loop gravity is derived from full loop quantum gravity by implementing a kind of quantum gauge-fixing to a gauge in which the densitized triad is diagonal, and the Hilbert space of the model is built entirely out of states based on cubical graphs.

In the context of quantum-reduced loop gravity, the new curvature operator of \cite{part1} represents a definite improvement over the operator introduced earlier in \cite{curvature}, where the basic ideas of Regge calculus were used to quantize the Ricci scalar in terms of lengths and angles. In general, operators for quantum-reduced loop gravity can be derived from the corresponding operators of full loop quantum gravity by applying the operators of the full theory on states in the Hilbert space of the quantum-reduced loop gravity and discarding certain small terms, which generally lie outside of the Hilbert space of the quantum-reduced model, in the resulting expressions \cite{IM}. However, if this procedure is applied to the curvature operator of \cite{curvature}, one finds that the action of the resulting operator is trivially vanishing on the reduced Hilbert space. In contrast, the new operator of \cite{part1} does give rise to a non-trivial curvature operator for the quantum-reduced model.

In this article we study the curvature operator introduced in \cite{part1} in the framework of quantum-reduced loop gravity. The main part of the work consists of performing the calculations required to establish the form of the curvature operator as an operator on the Hilbert space of the quantum-reduced model. We also examine the action of the resulting operator on the basis states of the reduced Hilbert space and its expectation values with respect to these states in order to gain a rudimentary picture of the basic properties of the operator. The material in the article is organized as follows. In section 2 be briefly recall the kinematical setting of quantum-reduced loop gravity, \ie the reduced Hilbert space and the elementary operators thereon. In section 3 we review the definition of the curvature operator given in our earlier article \cite{part1}. In section 4, the form of the curvature operator as an operator of quantum-reduced loop gravity is derived by computing the action of the operator on states in the reduced Hilbert space. In section 5 we discuss the computation of expectation values of the curvature operator in reduced basis states. Finally, we summarize and discuss our results in the concluding section 6. In the two appendixes that are included in the article, we recall a few useful facts from the quantum theory of angular momentum, and present some of the more technical segments of our calculations.

\section{Quantum-reduced loop gravity}
\label{sec:qrlg}

\subsection{The reduced Hilbert space}
\label{sec:H_red}

In this article we consider the Hilbert space of quantum-reduced loop gravity in the form originally introduced in the literature of the model \cite{qrlg1, qrlg2}. This Hilbert space is a proper subspace of the kinematical Hilbert space of loop quantum gravity. The Hilbert space of the quantum-reduced model is spanned by the so-called reduced spin network states, \ie states characterized by the following assumptions:
\begin{itemize}
	\item The state is based on a cubical graph, denoted by $\Gc$, whose edges are aligned along the coordinate directions defined by a fixed Cartesian background coordinate system.
	\item Each edge of the graph carries a large spin quantum number:
	\begin{equation}
		j_e \gg 1
		\label{j>>1}
	\end{equation}
	for every edge $e\in\Gc$.
	\item To each edge is assigned a representation matrix, both of whose magnetic indices take the maximal or the minimal value (\ie $+j_e$ or $-j_e$) with respect to the direction of the edge (in the sense explained by \Eqs{D_i} and \eqref{basis_r} below).
\end{itemize}

Let us introduce the notation $\ket{jm}_i$ (where $i = x$, $y$ or $z$) for the eigenstate of the operators $J^2$ and $J_i$ with eigenvalues $j(j+1)$ and $m$, and
\begin{equation}
	D^{(j)}_{mn}(h)_i = {}_i\bra{jm}D^{(j)}(h)\ket{jn}_i
	\label{D_i}
\end{equation}
for the matrix elements of the Wigner matrices with respect to the basis $\ket{jm}_i$. The reduced spin network states, which form a basis on the Hilbert space of quantum-reduced loop gravity, are then defined by wave functions of the form
\begin{equation}
	\prod_{e\in\Gc} D^{(j_e)}_{\sigma_ej_e\;\sigma_ej_e}(h_e)_{i_e}
	\label{basis_r}
\end{equation}
where each $\sigma_e$ takes the value $+1$ or $-1$, and each $i_e = x$, $y$ or $z$, according to whether the edge $e$ is aligned along the direction of the $x$-, $y$- or $z$-axis.

Note that the relation
\begin{equation}
	D^{(j)}_{jj}(h^{-1}) = D^{(j)}_{-j\;-j}(h)
	\label{}
\end{equation}
can be used to slightly simplify the bookkeeping associated with the basis states \eqref{basis_r}. If one keeps track of the orientation of the graph on which the state \eqref{basis_r} is defined, considering different orientations of the graph to define different, inequivalent states, one can without loss of generality set $\sigma_e = +1$ for every edge of the graph. Alternatively, one can take the graph $\Gc$ with an arbitrary but fixed orientation, while allowing the parameter $\sigma_e$ to take the value $+1$ or $-1$ independently for each edge of the graph.

\subsection{Operators in the quantum-reduced model}
\label{sec:red-operators}

The form of the elementary operators in quantum-reduced loop gravity can be derived by applying the corresponding operators of full loop quantum gravity on the basis states \eqref{basis_r}. The approximation inherent to the quantum-reduced model then amounts to discarding certain terms which are of lower order in the spin quantum numbers, and which can be neglected on grounds of the assumption \eqref{j>>1}, even if these terms generally do not belong to the reduced Hilbert space introduced in the previous section \cite{IM}. The structure of the kinematical operators obtained in this way is remarkably simple in comparison with the full theory, which is an important practical advantage of the quantum-reduced model from the point of view of performing concrete calculations.

The elementary operators typically considered in quantum-reduced loop gravity are holonomy operators associated to the edges of the cubical graph, and flux operators associated to surfaces which are dual to the background coordinate directions (\ie surfaces $S^a$ such that the coordinate $x^a = {\rm const.}$ on the surface) and are located at the nodes of the reduced spin network graph. The action of the holonomy operator $D^{(s)}_{mn}(h_e)_i$ on the basis state \eqref{basis_r} can be computed using the standard Clebsch--Gordan series of $SU(2)$. In order to express the result of the calculation, it is convenient to introduce the notation
\begin{equation}
	\DD^{(j)}_{mn}(h) = \sqrt{2j+1}\,D^{(j)}_{mn}(h)
	\label{}
\end{equation}
for the normalized\footnote{In the sense of the norm defined by the Haar measure on $SU(2)$.} matrix elements of the $SU(2)$ representation matrices. Then, under the assumption that the spin carried by the operator is small in comparison with the spin carried by the state, \ie $s \ll j$, one finds
\begin{align}
	D^{(s)}_{mm}(h_e)_i\DD^{(j)}_{jj}(h_e)_i &= \DD^{(j+m)}_{j+m\; j+m}(h_e)_i + {\cal O}\biggl(\frac{1}{j}\biggr) \label{Dmm_r} \\
\intertext{and}
	D^{(s)}_{mn}(h_e)_i\DD^{(j)}_{jj}(h_e)_i &= {\cal O}\biggl(\frac{1}{\sqrt j}\biggr) \qquad (m\neq n) \label{Dmn_r}
\end{align}
In other words, only the diagonal matrix elements contribute to the action of the holonomy operator at leading order in $j$. The leading term has essentially the structure of a $U(1)$ multiplication law, with the magnetic index (and not the spin) of the operator $D^{(s)}_{mm}(h_e)_i$ playing the role of the $U(1)$ quantum number. A detailed presentation of the calculations leading up to \Eqs{Dmm_r} and \eqref{Dmn_r} (as well as \Eqs{E_r} and \eqref{V_r} below), including a more thorough discussion of the discarded lower-order terms, can be found in \cite{IM}.

Consider then the action of the flux operator $E_i(S^a)$ on a ''reduced holonomy'' of the form $D^{(j)}_{jj}(h_e)_k$. The action can be non-trivial only in the case $a = k$, since otherwise the direction of the edge is parallel to the surface $S^a$. Assuming for concreteness that the surface intersects the edge at its beginning point\footnote{
	A similar calculation holds for the case where the edge $e$ is intersected by the surface $S^k$ at its endpoint. Note, however, that if the surface intersects the edge at an interior point, the action of the flux operator gives
\[
	E_i(S^k)D^{(j)}_{jj}(h_e)_k = i\Bigl(D^{(j)}(h_{e_1})\tau_i D^{(j)}(h_{e_2})\Bigr)_{jj}^{(k)}
\]
which involves all the matrix elements of the generator $\tau_i^{(j)}$, and which generally does not reduced to any simple form even for large $j$.
}, we have
\begin{equation}
	E_i(S^k)D^{(j)}_{jj}(h_e)_k = \frac{i}{2}\Bigl(D^{(j)}(h_e)\tau_i^{(j)}\Bigr)_{jj}^{(k)}
	\label{}
\end{equation}
with the superscript indicating that the matrix element is taken in the basis in which $J_k$ is diagonal. In this basis the diagonal generator $\tau_k^{(j)}$ has the matrix element $(\tau_k^{(j)})_{jj} = -ij$, whereas the relevant matrix elements of the off-diagonal generators $\tau_i^{(j)}$ $(i\neq k)$ are of order $\sqrt j$. Therefore we conclude that
\begin{equation}
	E_i(S^k)\DD^{(j)}_{jj}(h_e)_k = \begin{cases}
		\dfrac{j}{2}\DD^{(j)}_{jj}(h_e)_k & (i = k) \\[2ex]
		\Oj & (i\neq k)
	\end{cases}
	\label{E_r}
\end{equation}
The general structure of this result is quite similar to the action of the holonomy operator given by \Eqs{Dmm_r} and \eqref{Dmn_r}, with the contribution of leading order in $j$ again given by the diagonal (\ie $i = k$) component of the operator.

\begin{figure}[t]
	\centering
	\includegraphics[scale=0.16]{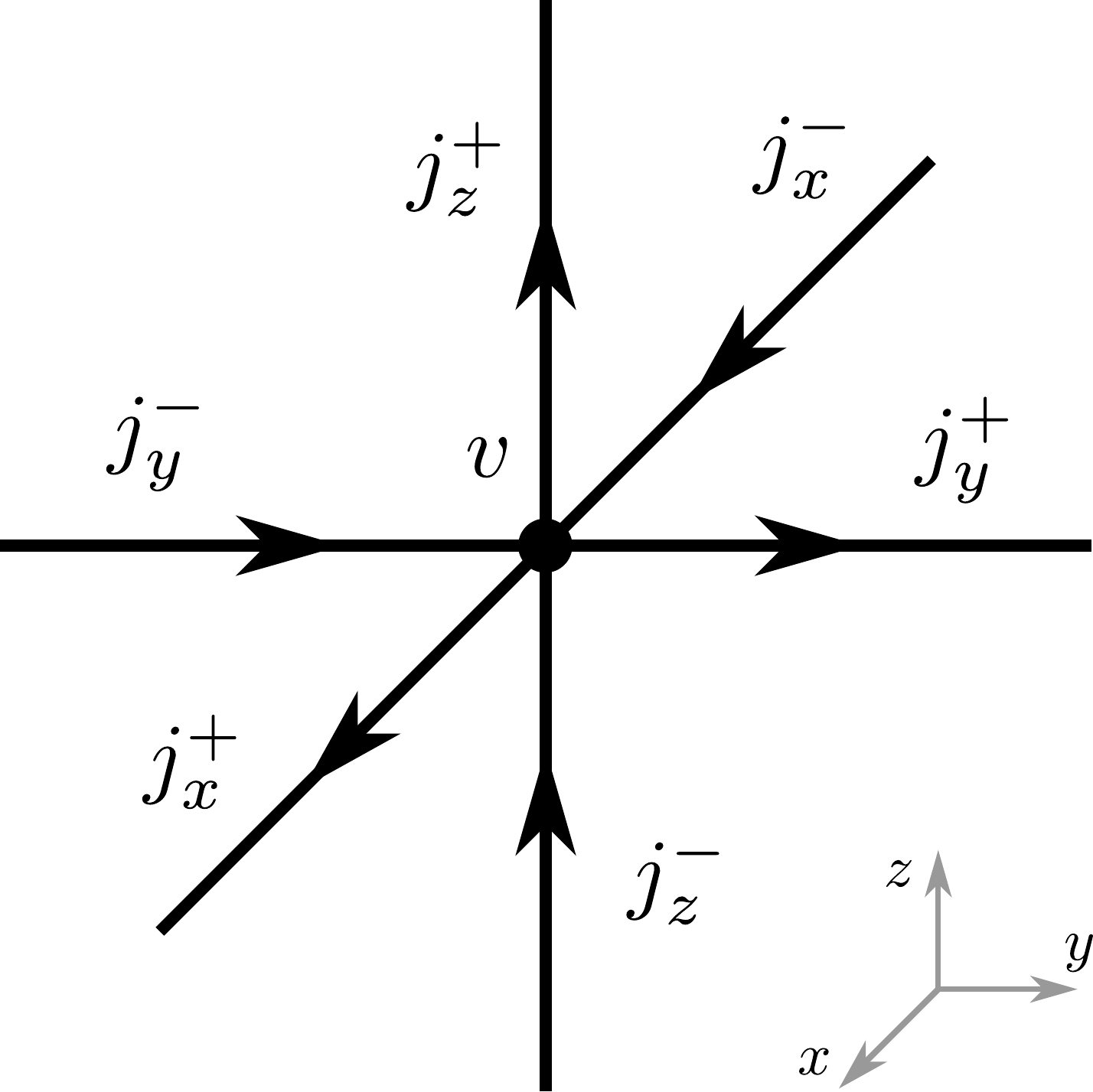}
	\caption{A generic node of a reduced spin network state.}
	\label{fig:node}
\end{figure}

The simple, approximately diagonal form of the flux operator given by \Eq{E_r} is naturally inherited by operators constructed out of the flux operator. An important example of such an operator is the volume operator \cite{volume}. When restricted to a six-valent node of a cubical graph, the Ashtekar--Lewandowski volume operator can be expressed in terms of flux operators in the form
\begin{equation}
	V_v = \sqrt{|W_v|}
	\label{V_v}
\end{equation}
with
\begin{equation}
	W_v = \epsilon^{ijk}E_i\bigl(S^x(v)\bigr)E_j\bigl(S^y(v)\bigr)E_k\bigl(S^z(v)\bigr).
	\label{W_v}
\end{equation}
Here $S^x(v)$, $S^y(v)$ and $S^z(v)$ are surfaces which contain the node $v$ (while not containing any other nodes of the graph), and which and are dual to the corresponding background coordinate directions. Using \Eq{E_r} to compute the action of the operator \eqref{W_v} on a reduced spin network state, one finds a diagonal leading term of order $j^3$, together with subleading, off-diagonal terms of order $j^2$. The separation between the leading and subleading terms is preserved by the operation of taking the square root in \Eq{V_v} \cite{IM}. With the spins labeled as in \Fig{fig:node}, the result of the calculation reads
\begin{equation}
	V_v\ket{\Psi_\Gc} = \sqrt{\frac{1}{8}\Bigl|\bigl(j_x^+ + j_x^-\bigr)\bigl(j_y^+ + j_y^-\bigr)\bigl(j_z^+ + j_z^-\bigr)\Bigr|}\,\ket{\Psi_\Gc}  + {\cal O}\bigl(\sqrt j\bigr).
	\label{V_r}
\end{equation}
Clearly this represents a considerable simplification in comparison with the task of computing the action of the volume operator on a generic spin network state in full loop quantum gravity.

\subsection{Notation for reduced operators}
\label{sec:notation}

The essential approximation underlying quantum-reduced loop gravity consists of discarding the terms of lower order in $j$ in \Eqs{Dmm_r}, \eqref{Dmn_r} and \eqref{E_r}. This step can be justified by the assumption \eqref{j>>1}, which specifies that the states in the reduced Hilbert space carry large spins on all of their edges. The action of the operators obtained as a result of such a truncation preserves the reduced Hilbert space. These operators, which are commonly known as reduced holonomy and flux operators in the literature of the quantum-reduced model, are therefore well-defined operators on the reduced Hilbert space.

As a preparation for the calculations performed in the main part of this work, we will now establish a condensed notation for the reduced holonomy and flux operators. The components of the reduced holonomy operator, which correspond to the diagonal matrix elements $D^{(s)}_{mm}(h_{e_i})_i$ of the full holonomy operator, will be denoted by $d_m(e_i)$, with the subscript $i$ indicating the direction of the edge $e_i$. The action of the operator $d_m(e_i)$ on the reduced Hilbert space is defined by
\begin{equation}
	d_m(e_i)D^{(j)}_{jj}(h_{e_i})_i = D^{(j+m)}_{j+m\; j+m}(h_{e_i})_i.
	\label{}
\end{equation}
In the particular case $j=1$, which will feature prominently in our calculations, we denote the three possible values of the index $m$ by $+$, $0$ and $-$, thus introducing the notation
\begin{equation}
	d_+(e_i), \qquad d_0(e_i) = \Id(e_i), \qquad d_-(e_i)
	\label{}
\end{equation}
for the reduced holonomy operator in the $j=1$ representation.

The reduced flux operator arises from the diagonal ($i=a$) component of the flux operator $E_i\bigl(S^a(v)\bigr)$, where the surface $S^a(v)$ is dual to the $x^a$ -coordinate direction and contains the node $v$ of a reduced spin network state. We denote this operator as
\begin{equation}
	p_a(v) = E_a\bigl(S^a(v)\bigr) 
	\label{}
\end{equation}
The action of the reduced flux operator $p_a(v)$ on a reduced spin network state is diagonal, with the eigenvalue
\begin{equation}
	p_a(v) = \frac{j_a^+(v) + j_a^-(v)}{2},
	\label{}
\end{equation}
where each of the two edges incident on the node $v$ and parallel to the $x^a$-coordinate axis contributes to the eigenvalue according to \Eq{E_r}. The reduced volume operator, which is obtained by keeping only the leading term in \Eq{V_r}, can be expressed in terms of reduced flux operators as
\begin{equation}
	\sqrt{|w(v)|}
	\label{}
\end{equation}
where
\begin{equation}
	w(v) = p_x(v)p_y(v)p_z(v).
	\label{}
\end{equation}

\section{The curvature operator}
\label{sec:R}

In this section we summarize the operator introduced in \cite{part1}, which represents the scalar curvature of the spatial manifold, and which is defined on the Hilbert space of a fixed cubical graph. The operator is obtained as the result of a construction which begins by expressing the three-dimensional Ricci scalar as a function of the densitized triad $E^a_i$ and its gauge covariant derivatives, $\D_aE^b_i = \partial_aE^b_i + \epsilon\downup{ij}{k}A_a^jE^b_k$. The integrated Ricci scalar $\int d^3x\,\sqrt q\Rt$ is then regularized in terms of a cellular decomposition of the spatial manifold into cubical cells adapted to the chosen cubical graph. The elementary operators entering the definition of the resulting curvature operator are holonomy operators associated to the edges of the cubical graph, and flux operators associated to the surfaces $S^x(v)$, $S^y(v)$ and $S^z(v)$, which are located at the nodes of the cubical graph, and which are dual to the coordinate directions defined by the background coordinate system, along which the edges of the cubical graph are aligned.

On the Hilbert space of the cubical graph $\Gc$, the curvature operator takes the form
\begin{equation}
	\biggl(\widehat{\int d^3x\,\sqrt q\Rt}\biggr) = \sum_{v\in\Gc} \frac{{\cal R}_v}{V_v}.
	\label{R^}
\end{equation}
The operator ${\cal R}_v$ is defined as
\begin{align}
	{\cal R}_v = & -2\Et_i\bigl(S^a(v), v\bigr)\Delta_{ab}E_i\bigl(S^b, v\bigr) + 2{\cal Q}^{ab}(v)\widetilde{\cal E}_c^i\bigl(S(v), v\bigr)\Delta_{ab}E_i\bigl(S^c, v\bigr) \notag \\
	& -\Delta_aE_i\bigl(S^a, v\bigr)\Delta_bE_i\bigl(S^b, v\bigr) - \frac{1}{2}\Delta_aE_i\bigl(S^b, v\bigr)\Delta_bE_i\bigl(S^a, v\bigr) \notag \\
	& + \frac{5}{2}{\cal Q}^{ab}(v)\Delta_aE_i\bigl(S^c, v\bigr)\Delta_b{\cal E}_c^i(v) - \frac{1}{2}{\cal Q}^{ab}(v){\cal Q}_{cd}(v) \Delta_aE_i\bigl(S^c, v\bigr)\Delta_bE_i\bigl(S^d, v\bigr) \notag \\
	& + 2{\cal A}\updown{ab}{a}(v){\cal B}\downup{cb}{c}(v) + 2{\cal A}\updown{ab}{b}(v){\cal B}\downup{ca}{c}(v) + {\cal A}\updown{ab}{c}(v){\cal B}\downup{ba}{c}(v) \notag \\
	& + \frac{1}{2}{\cal Q}_{ab}(v){\cal A}\updown{ca}{d}(v){\cal A}\updown{db}{c}(v) - {\cal Q}^{ab}(v){\cal B}\downup{ca}{c}(v){\cal B}\downup{db}{d}(v) \notag \\
	& + 2\bigl({\cal Q}^{ab}(v){\cal B}\downup{ca}{c}(v) - {\cal A}\updown{ab}{a}(v) - {\cal A}\updown{ba}{a}(v)\bigr)\frac{\Delta_b V(v)^2}{V_v^2} \notag \\
	& + \frac{3}{2}{\cal Q}^{ab}(v)\frac{\Delta_a V(v)^2}{V_v^2}\frac{\Delta_b V(v)^2}{V_v^2} - 2{\cal Q}^{ab}(v)\frac{\Delta_{ab}V(v)^2}{V_v^2}.
	\label{R(v)}
\end{align}
where
\begin{align}
	{\cal Q}^{ab}(v) &= \Et_i\bigl(S^a(v), v\bigr)\Et_i\bigl(S^b(v), v\bigr) \label{QQ^ab} \\
	{\cal Q}_{ab}(v) &= \widetilde{\cal E}_a^i\bigl(S(v), v\bigr)\widetilde{\cal E}_b^i\bigl(S(v), v\bigr) \label{QQ_ab} \\
	{\cal A}\updown{ab}{c}(v) &= \Et_i\bigl(S^a(v), v\bigr)\Delta_c E_i\bigl(S^b, v\bigr) \\
	{\cal B}\downup{ab}{c}(v) &= \widetilde{\cal E}_a^i\bigl(S(v), v\bigr)\Delta_b E_i\bigl(S^c, v\bigr)
	\label{}
\end{align}
The various operators entering the definition of the curvature operator will be introduced below. For a detailed presentation of the construction we refer the reader to \cite{part1}.

\subsubsection*{Inverse volume operator}

The factors of volume operator in the denominator in \Eqs{R^} and \eqref{R(v)} are to be understood in terms of the regularized inverse volume operator ${\cal V}_v^{-1}$, which can be defined as the limit
\begin{equation}
	{\cal V}_v^{-1} = \lim_{\epsilon\to 0} \frac{V_v}{V_v^2 + \epsilon^2},
	\label{}
\end{equation}
where $V_v$ is the volume operator restricted to the node $v$. Equivalently, the operator ${\cal V}_v^{-1}$ can be defined by specifying its spectral decomposition as
\begin{equation}
	{\cal V}^{-1}_v\ket\lambda = \begin{cases} 
		\lambda^{-1}\ket\lambda & \text{if $\lambda\neq 0$} \\[0.5ex] 
		0 & \text{if $\lambda = 0$}
	\end{cases}
	\label{V_v^-1}
\end{equation}
where $\ket{\lambda}$ is an eigenstate of the volume operator $V_v$ with eigenvalue $\lambda$. The definition extends straightforwardly to any negative power of the volume: ${\cal V}_v^{-n} \equiv \bigl({\cal V}_v^{-1}\bigr)^n$ ($n > 0$).

The volume operator $V_v$ and the operator ${\cal R}_v$ defined by \Eq{R(v)} do not commute with each other, so a choice of factor ordering has to be made on the right-hand side of \Eq{R^}. The expression ${\cal R}_v/V_v$ should therefore be understood as a shorthand notation for any symmetric factor ordering of the operators ${\cal R}_v$ and ${\cal V}_v^{-1}$.

\subsubsection*{Parallel transported flux operator}

The parallel transported flux operator (also known as the gauge covariant flux operator in the literature) is a quantization of the classical function
\begin{equation}
	\Et_i(S, x_0) = -2\,{\rm Tr}\,\Bigl(\tau_i \Et(S, x_0)\Bigr)
	\label{Et_i}
\end{equation}
where $\Et(S, x_0)$ is the matrix-valued variable 
\begin{equation}
	\Et(S, x_0) = \int_S d^2\sigma\,n_a(\sigma)h_{x_0, x(\sigma)}E^a_i\bigl(x(\sigma)\bigr)\tau_i h^{-1}_{x_0, x(\sigma)}
	\label{Et}
\end{equation}
and $h_{x_0, x(\sigma)} \equiv h_{p_{x_0, x(\sigma)}}$ denote holonomies in the fundamental representation of $SU(2)$ (\ie $h_{x_0, x(\sigma)} \equiv D^{(1/2)}(h_{x_0, x(\sigma)})$). These holonomies connect each point $x(\sigma)$ on the surface $S$ to a fixed point $x_0$ along a family of paths $p_{x_0, x(\sigma)}$. Assuming there is a single point of intersection, denoted $v$, between an edge $e$ and the surface $S$, the action of the parallel transported flux operator on the holonomy $h_e$ can be expressed as
\begin{equation}
	\Et_i(S, x_0)D^{(j)}(h_e) = D^{(1)}_{ki}\bigl(h_{x_0, v}^{-1}\bigr) E_k(S)D^{(j)}(h_e),
	\label{Et-action}
\end{equation}
where $E_k(S)$ denotes the standard flux operator associated to the surface $S$, and the holonomy operator in the spin-1 representation has arisen from the action of the holonomies in the fundamental representation on the generator $\tau_i$ according to the identity $h\tau_i h^{-1} = D^{(1)}_{ki}(k)\tau_k$. A more detailed presentation of the material summarized above can be found \eg in section III.C of \cite{part1}.

\subsubsection*{Inverse flux operator}

Using the parallel transported flux operator, we define
\begin{equation}
	\EEt_a^i\bigl(S(v), v'\bigr) = \frac{1}{2}\epsilon_{abc}\epsilon^{ijk}\Et_j\bigl(S^b(v), v'\bigr)\Et_k\bigl(S^c(v), v'\bigr){\cal W}_v^{-1}.
	\label{EEt}
\end{equation}
Here ${\cal W}_v^{-1}$ denotes a regularized inverse of the operator
\begin{equation}
	W_v = \epsilon^{ijk}E_i\bigl(S^x(v)\bigr)E_j\bigl(S^y(v)\bigr)E_k\bigl(S^z(v)\bigr),
	\label{}
\end{equation}
defined in a way analogous to \Eq{V_v^-1}, namely
\begin{equation}
	{\cal W}^{-1}_v\ket\mu = \begin{cases} 
		\lambda^{-1}\ket\mu & \text{if $\mu\neq 0$} \\[0.5ex] 
		0 & \text{if $\mu = 0$}
	\end{cases}
	\label{W_v^-1}
\end{equation}
where $\ket\mu$ is an eigenstate of the operator $W_v$ with eigenvalue $\mu$.

When $v' = v$, the operator \eqref{EEt} represents a quantization of the inverse triad
\begin{equation}
	E_a^i = \frac{1}{2\det E}\epsilon_{abc}\epsilon^{ijk}E^b_jE^c_k
	\label{}
\end{equation}
at the node $v$ -- in the same, not entirely precise sense in which the flux operator $E_i\bigl(S^a(v)\bigr)$ represents a quantization of the densitized triad $E^a_i(v)$ itself.

\subsubsection*{Discretized derivatives}

The role of the parallel transported flux operator in the construction of the curvature operator is to provide a tool for quantizing covariant derivatives of the densitized triad. The operator $\Delta_aE_i\bigl(S^b, v\bigr)$, which arises from a discretization of the covariant derivative $\D_aE^b_i$ at the point $v$, is defined as
\begin{equation}
	\Delta_aE_i\bigl(S^b, v\bigr) = \frac{\Et_i\bigl(S^b(v_a^+), v\bigr) - \Et_i\bigl(S^b(v_a^-), v\bigr)}{2}.
	\label{DaEb}
\end{equation}
Here $v_a^+$ and $v_a^-$ denote the nodes immediately following and immediately preceding the central node $v$ in the direction of the $x^a$ -coordinate axis. The parallel transport from $v_a^+$ and $v_a^-$ to $v$ is taken along the edges $e_a^+$ and $e_a^-$ connecting $v_a^+$ and $v_a^-$ to the central node $v$.

The operator
\begin{equation}
	\Delta_{aa}E_i\bigl(S^b, v\bigr) = \Et_i\bigl(S^b(v_a^+), v\bigr) - 2\Et_i\bigl(S^b(v), v\bigr) + \Et_i\bigl(S^b(v_a^-), v\bigr)
	\label{DaaEb}
\end{equation}
represents a quantization of the ''diagonal'' second derivative $\D_a^2E^b_i$ at $v$. Here two of the parallel transported flux operators are the same as in \Eq{DaEb}, and the action of the operator $\Et_i\bigl(S^b(v), v\bigr)$ at the central node is identical to the regular flux operator $E_i\bigl(S^b(v)\bigr)$.

\begin{figure}[t]
	\centering
	\includegraphics[scale=0.17]{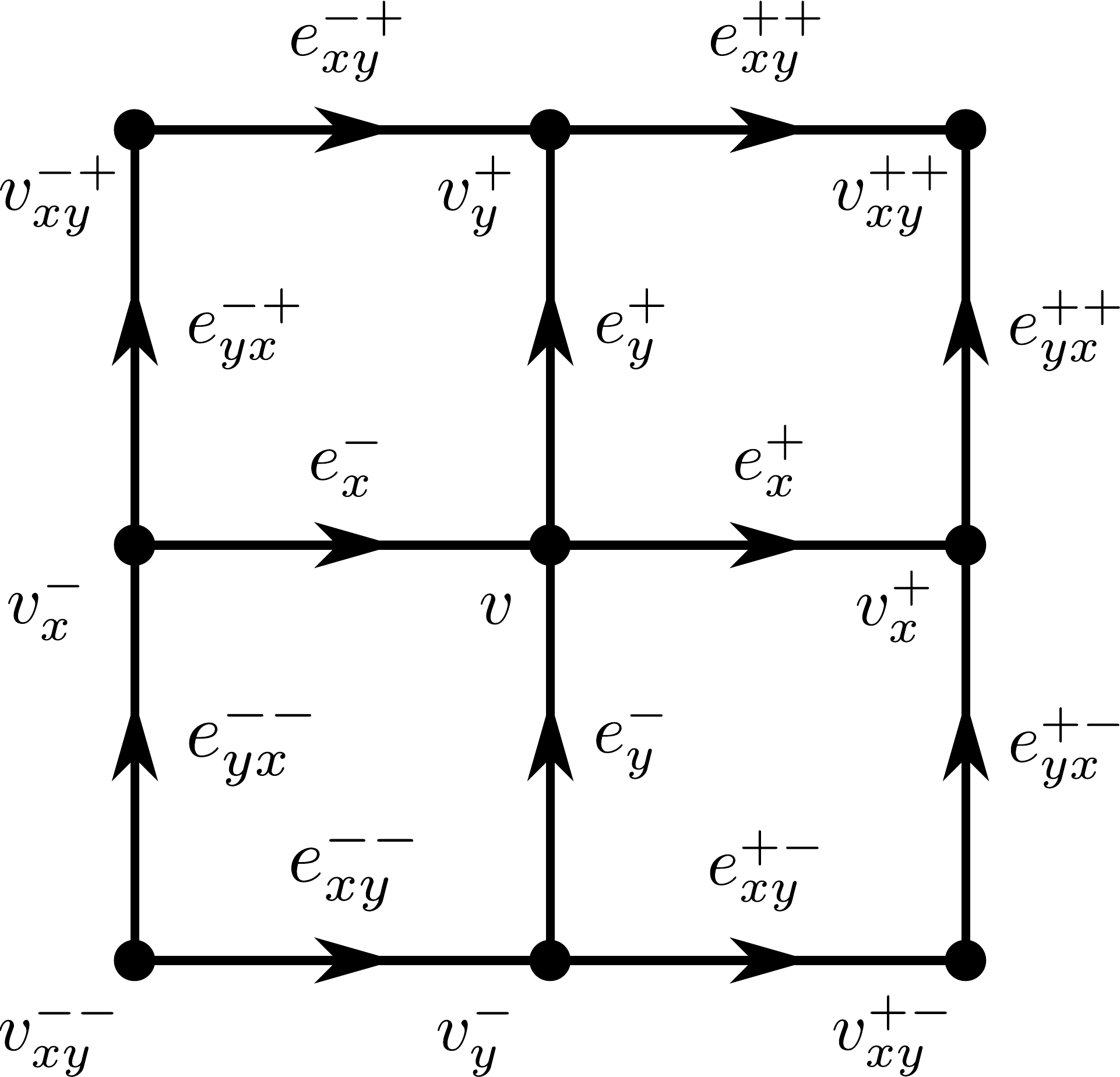}
	\caption{Labeling of the nodes and edges involved in the definition of the mixed components of the second derivative operator $\Delta_{ab}E_i\bigl(S^c, v\bigr)$.}
	\label{labels}
\end{figure}

For $a\neq b$, the setup and notation\footnote{
	The pattern behind the notation for the edges entering the definition is as follows: The edge which lies in the $(x^a, x^b)$ -coordinate plane, is connected to the node $v_{ab}^\sigma$ and oriented along the $x^a$ -coordinate axis is denoted by $e_{ab}^\sigma$. Thus, $e_{ba}^\sigma$ denotes the edge lying in the $(x^a, x^b)$-plane and connected to $v_{ab}^\sigma$, but oriented in the direction of the $x^b$-axis.
} for defining the operator $\Delta_{ab}E_i\bigl(S^b, v\bigr)$ is given in \Fig{labels}. The discretization of the mixed second derivative $\D_a\D_bE^c_i(v)$ is performed using the four nodes diagonally adjacent to the central node $v$ in the $(x^a, x^b)$ -coordinate plane. Let us denote these nodes by $v_{ab}^\sigma$, where the symbol
\begin{equation}
	\sigma = \bigl(\sigma^1, \sigma^2\bigr) = (++), (+-), (-+), (--)
	\label{}
\end{equation}
labels the four quadrants of the $(x^a, x^b)$ -coordinate plane. We then define the symmetrized flux operators
\begin{equation}
	\Et_i\bigl(S^c(v_{ab}^{\sigma}), v\bigr)_{\rm sym.} = \frac{1}{2}\Bigl(\Et_i\bigl(S^c(v_{ab}^{\sigma}), v\bigr)_{v_{ab}^{\sigma} \to v_a^{\sigma^1} \to v} + \Et_i\bigl(S^c(v_{ab}^{\sigma}), v\bigr)_{v_{ab}^{\sigma} \to v_b^{\sigma^2} \to v}\Bigr),
	\label{E_sym}
\end{equation}
where the subscripts on the right-hand side refer to the two natural routes along which the parallel transport from $v_{ab}^\sigma$ to $v$ can be taken. The operator $\Delta_{ab}E_i\bigl(S^c, v\bigr)$ (for $a\neq b$) is now defined as
\begin{align}
	\Delta_{ab}E_i\bigl(S^c, v\bigr) = \frac{1}{4}\biggl(\Et_i\bigl(&S^c(v_{ab}^{++}), v\bigr)_{\rm sym.} - \Et_i\bigl(S^c(v_{ab}^{+-}), v\bigr)_{\rm sym.} \notag \\
	&- \Et_i\bigl(S^c(v_{ab}^{-+}), v\bigr)_{\rm sym.} + \Et_i\bigl(S^c(v_{ab}^{--}), v\bigr)_{\rm sym.}\biggr).
	\label{DabEc}
\end{align}
By construction, the operator \eqref{DabEc} is symmetric in $a$ and $b$. Accordingly, it represents a quantization of the symmetric part $\D_{(a}\D_{b)}E^c_i$ of the mixed second derivative at $v$.

\Eq{R(v)} also features the operator $\Delta_a\EE_b^i(v)$, which corresponds to a quantization of the covariant derivative of the inverse triad $E_a^i$. Making use of the operator \eqref{EEt}, this operator is defined as
\begin{equation}
	\Delta_a\EE_b^i(v) = \frac{\threedots\EEt_b^i\bigl(S(v_a^+), v\bigr)\threedots - \threedots\EEt_b^i\bigl(S(v_a^-), v\bigr)\threedots}{2}.
	\label{DaEEb}
\end{equation}
The notation with three dots indicates a particular factor ordering of the operators, in which the holonomy operators arising from the parallel transported flux operators in \Eq{EEt} are ordered as the leftmost factor of the expression, as will be specified in detail in section \ref{sec:inverse}.

Finally, the discretized derivatives of the volume are defined by the expressions
\begin{equation}
	\Delta_a V(v)^2 = \frac{V(v_a^+)^2 - V(v_a^-)^2}{2},
	\label{ddet}
\end{equation}
\begin{equation}
	\Delta_{aa} V(v)^2 = V(v_a^+)^2 - 2V(v)^2 + V(v_a^-)^2
	\label{d2det}
\end{equation}
and
\begin{equation}
	\Delta_{ab} V(v)^2 = \frac{V(v_{ab}^{++})^2 - V(v_{ab}^{+-})^2 - V(v_{ab}^{-+})^2 + V(v_{ab}^{--})^2}{4}
	\label{dddet}
\end{equation}
All of the operators \eqref{ddet}--\eqref{dddet} commute with the volume operator $V_v \equiv V(v)$, so there is no ordering ambiguity between these operators and the inverse volume operators in \Eq{R(v)}.

This completes the definition of the curvature operator, with \Eqs{R^} and \eqref{R(v)} defining the action of the operator on any state which is based on the chosen cubical graph $\Gc$, but which may otherwise be a completely general state in the kinematical Hilbert space of loop quantum gravity. In what follows, we will consider the action of this operator on reduced spin network states in the Hilbert space of quantum-reduced loop gravity, thereby deriving the form of the curvature operator for the quantum-reduced model. Even though the calculations are somewhat lengthy, they provide a practical illustration of the procedure of extracting an operator for the quantum-reduced model from the corresponding operator of the full theory, and for this reason we feel that it is valuable to present these calculations in some detail.

\section{Curvature operator on the reduced Hilbert space}
\label{sec:R-reduced}

In this section we will derive the form of the curvature operator when it is interpreted as an operator on the Hilbert space of quantum-reduced loop gravity. To this end, we will consider the action of the curvature operator on the reduced spin network states, which form a basis of the Hilbert space of the quantum-reduced model. The reduced spin network states are based on a cubical graph, whose edges are aligned with the coordinate directions defined by a fixed Cartesian background coordinate system. We assume that the orientation of each edge of the graph agrees with the positive direction of the corresponding coordinate axis, and that each edge carries a ''reduced holonomy'' of the form
\begin{equation}
	D^{(j_e)}_{j_ej_e}(h_e)_{i_e},
	\label{}
\end{equation}
where the label $i_e$ takes the value $x$, $y$ or $z$ according to the direction of the edge $e$.

To obtain the curvature operator as an operator on the reduced Hilbert space, we must evaluate the action of the operator on states of this form and truncate the resulting expressions at leading order in the spin quantum numbers. As discussed in section \ref{sec:qrlg}, the leading terms will arise from the diagonal $(m=n)$ components of the holonomy operators $D^{(j)}_{mn}(h_e)_{i_e}$, and from the diagonal $(i=a)$ components of the flux operators $E_i\bigl(S^a(v)\bigr)$. The terms of leading order in $j$ are guaranteed to belong to the reduced Hilbert space, while the lower-order terms, which are discarded, lie outside of this space in general. Hence the action of the operator truncated in this way will preserve the reduced Hilbert space, and the truncated operator will be a well-defined operator on this space.

Let us focus on the operator ${\cal R}_v$ defined by \Eq{R(v)}, and start by considering the factors of $\Et_i\bigl(S^a(v), v\bigr)$ and $\EEt_a^i\bigl(S(v), v\bigr)$. The action of the parallel transported flux operator $\Et_i\bigl(S^a(v), v\bigr)$ coincides with the standard flux operator $E_i\bigl(S^a(v)\bigr)$, and when applied on the reduced Hilbert space, the diagonal components of this operator form the reduced flux operator $p_a(v)$ introduced in section \ref{sec:notation}. Moreover, a short calculation shows that the leading terms in the action of the operator $\EEt_a^i\bigl(S(v), v\bigr)$ on a reduced spin network state also arise from the terms with $i=a$, and that these terms act as the inverse of the operator $p_a(v)$. Thus, making the replacements
\begin{align}
	\Et_i\bigl(S^a(v), v\bigr) &\rightarrow \delta^a_i p_a(v), \\
	\EEt_a^i\bigl(S(v), v\bigr) &\rightarrow \delta_a^i\frac{1}{p_a(v)}
	\label{}
\end{align}
in \Eq{R(v)}, we arrive at
\newpage
\begin{align}
	{\cal R}_v = & -2p_a(v)\Delta_{ab}E_a\bigl(S^b, v\bigr) + 2\frac{p_a(v)^2}{p_b(v)}\Delta_{aa}E_b\bigl(S^b, v\bigr) \notag \\
	& -\Delta_aE_c\bigl(S^a, v\bigr)\Delta_bE_c\bigl(S^b, v\bigr) - \frac{1}{2}\Delta_aE_c\bigl(S^b, v\bigr)\Delta_bE_c\bigl(S^a, v\bigr) \notag \\
	& + \frac{5}{2}p_a(v)^2\Delta_aE_c\bigl(S^b, v\bigr)\Delta_a{\cal E}_b^c(v) - \frac{1}{2}\frac{p_a(v)^2}{p_b(v)^2}\Delta_aE_c\bigl(S^b, v\bigr)\Delta_aE_c\bigl(S^b, v\bigr) \notag \\
	& + 2p_a(v)\Delta_aE_a\bigl(S^b, v\bigr)\frac{1}{p_c(v)}\Delta_bE_c\bigl(S^c, v\bigr) + 2p_a(v)\Delta_bE_a\bigl(S^b, v\bigr)\frac{1}{p_c(v)}\Delta_aE_c\bigl(S^c, v\bigr) \notag \\
	& + p_a(v)\Delta_cE_a\bigl(S^b, v\bigr)\frac{1}{p_b(v)}\Delta_aE_b\bigl(S^c, v\bigr) + \frac{1}{2}\frac{p_b(v)}{p_a(v)^2}\Delta_cE_b\bigl(S^a, v\bigr)p_c(v)\Delta_bE_c\bigl(S^a, v\bigr) \notag \\
	& - \frac{p_a(v)^2}{p_b(v)}\Delta_aE_b\bigl(S^b, v\bigr)\frac{1}{p_c(v)}\Delta_aE_c\bigl(S^c, v\bigr) \notag \\
	& + 2\biggl(\frac{p_a(v)^2}{p_b(v)}\Delta_aE_b\bigl(S^b, v\bigr) - p_b(v)\Delta_bE_b\bigl(S^a, b\bigr) - p_a(v)\Delta_bE_a\bigl(S^b, v\bigr)\biggr)\frac{\Delta_a V(v)^2}{V(v)^2} \notag \\
	& + \frac{3}{2}p_a(v)^2\biggl(\frac{\Delta_a V(v)^2}{V(v)^2}\biggr)^2 - 2p_a(v)^2\frac{\Delta_{aa}V(v)^2}{V(v)^2}.
	\label{Rv-reduced}
\end{align}
Here the discretized derivatives of the volume, defined by \Eqs{ddet}--\eqref{dddet}, act diagonally on reduced spin network states. Hence the remaining non-trivial part of the calculation, to which we will now turn our attention, is to derive the form of the operators $\Delta_aE_i\bigl(S^b, v\bigr)$, $\Delta_{ab}E_i\bigl(S^c, v\bigr)$ and $\Delta_a{\cal E}_b^i(v)$ when these are viewed as operators on the reduced Hilbert space.

\subsection{First derivatives: $\Delta_aE_i\bigl(S^b, v\bigr)$}
\label{sec:first}

We begin by looking at the operator
\begin{equation}
	\Delta_aE_i\bigl(S^b, v\bigr) = \frac{\Et_i\bigl(S^b(v_a^+), v\bigr) - \Et_i\bigl(S^b(v_a^-), v\bigr)}{2}.
	\label{DaEb-2}
\end{equation}
Let us first fix $a=z$ and consider the various possible values of the labels $b$ and $i$. Using \Eq{Et-action}, we find that when acting on a reduced spin network state, each of the parallel transported fluxes in \Eq{DaEb-2} gives the contribution\footnote{Note that, due to the presence of the holonomy operator in \Eq{Et-contribution}, the action of the parallel transported flux operator $\Et_i\bigl(S^b(v_z^\pm), v\bigr)$ on a reduced spin network state can be non-vanishing even if $i\neq b$, in contrast to the regular flux operator $E_i\bigl(S^b(v_z^\pm)\bigr)$, where only the diagonal $(i=b)$ components contribute at leading order in $j$.}
\begin{equation}
	\Et_i\bigl(S^b(v_z^\pm), v\bigr) \to D_{bi}^{(1)}\bigl(h_{v, v_z^\pm}^{-1}\bigr)p_b(v_z^\pm),
	\label{Et-contribution}
\end{equation}
where $h_{v, v_z^\pm}$ denotes the holonomy along the edge $e_{v, v_z^\pm}$, which connects the node $v_a^\pm$ to $v$ and is oriented from the outer node $v_a^\pm$ towards the central node $v$. In terms of the labeling summarized in \Fig{labels}, we thus have
\begin{equation}
	e_{v, v_z^+} = (e_z^+)^{-1}
	\label{e+}
\end{equation}
and
\begin{equation}
	e_{v, v_z^-} = e_z^-.
	\label{e-}
\end{equation}

The matrix elements of the holonomy in \Eq{Et-contribution} must now be transformed from the Cartesian basis to the basis diagonalizing $J_z$, which is the component of the angular momentum operator corresponding to the direction of the edge $e_{v, v_z^\pm}$. In the eigenbasis of $J_z$, the leading order contribution will be given by the diagonal matrix elements of the holonomy operator, while the action of the off-diagonal matrix elements is of lower order in $j$. Applying the relations
\begin{align}
	\ket{x} &= -\frac{1}{\sqrt{2}}\bigl(\ket{+}_z - \ket{-}_z\bigr), \\
	\ket{y} &= \frac{i}{\sqrt{2}}\bigl(\ket{+}_z + \ket{-}_z\bigr), \\
	\ket{z} &= \ket{0}_z,
	\label{}
\end{align}
which are established in appendix \ref{sec:A}, to the matrix elements $D^{(1)}_{bi}(h)$ and discarding the off-diagonal matrix elements in the resulting expressions, we find
\begin{align}
	D_{xx}^{(1)}(h) &= \frac{1}{2}\Bigl(D^{(1)}_{11}(h_e)_z + D^{(1)}_{-1\; -1}(h)_z\Bigr) + \text{off-diag.} \label{D_xx} \\
	D_{xy}^{(1)}(h) &= -\frac{i}{2}\Bigl(D^{(1)}_{11}(h_e)_z - D^{(1)}_{-1\; -1}(h)_z\Bigr) + \text{off-diag.} \label{D_xy} \\
	D_{xz}^{(1)}(h) &= \text{off-diag.}
	\label{}
\end{align}
and
\begin{align}
	D_{yx}^{(1)}(h) &= \frac{i}{2}\Bigl(D^{(1)}_{11}(h)_z - D^{(1)}_{-1\; -1}(h)_z\Bigr) + \text{off-diag.} \label{D_yx} \\
	D_{yy}^{(1)}(h) &= \frac{1}{2}\Bigl(D^{(1)}_{11}(h)_z + D^{(1)}_{-1\; -1}(h)_z\Bigr) + \text{off-diag.} \label{D_yy} \\
	D_{yz}^{(1)}(h) &= \text{off-diag.}
	\label{}
\end{align}
and finally
\begin{align}
	D_{zx}^{(1)}(h_e) &= \text{off-diag.} \\
	D_{zy}^{(1)}(h_e) &= \text{off-diag.} \\
	D_{zz}^{(1)}(h_e) &= D_{00}^{(1)}(h_e)_z
	\label{}
\end{align}
When these results are used in \Eq{Et-contribution}, we must keep in mind the orientation of the edges \eqref{e+} and \eqref{e-} when identifying the diagonal matrix elements of the holonomy with the operators $d_\pm(e)$ introduced in section \ref{sec:notation}. For the edge $e_{v, v_z^+} = (e_z^+)^{-1}$ we have
\begin{align}
	D^{(1)}_{11}\bigl(h_{v, v_z^+}^{-1}\bigr) &\to d_+(e_z^+) \\
	D^{(1)}_{00}\bigl(h_{v, v_z^+}^{-1}\bigr) &\to \Id(e_z^+) \\
	D^{(1)}_{-1\; -1}\bigl(h_{v, v_z^+}^{-1}\bigr) &\to d_-(e_z^+)
	\label{}
\end{align}
whereas for $e_{v, v_z^-} = e_z^-$,
\begin{align}
	D^{(1)}_{11}\bigl(h_{v, v_z^-}^{-1}\bigr) &\to d_-(e_z^-) \\
	D^{(1)}_{00}\bigl(h_{v, v_z^-}^{-1}\bigr) &\to \Id(e_z^-) \\
	D^{(1)}_{-1\; -1}\bigl(h_{v, v_z^-}^{-1}\bigr) &\to d_+(e_z^-)
	\label{}
\end{align}
Going back to \Eq{DaEb-2}, we can now establish the form of the operator $\Delta_zE_i\bigl(S^b, v\bigr)$ as an operator on the reduced Hilbert space. The results are as follows. For $b=x$, we have
\begin{align}
	\Delta_zE_x\bigl(S^x, v\bigr) &= \frac{1}{4}\Bigl(d_+(e_z^+) + d_-(e_z^+)\Bigr)p_x(v_z^+) - \frac{1}{4}\Bigl(d_+(e_z^-) + d_-(e_z^-)\Bigr)p_x(v_z^-) \label{DE-zxx} \\
	\Delta_zE_y\bigl(S^x, v\bigr) &= -\frac{i}{4}\Bigl(d_+(e_z^+) - d_-(e_z^+)\Bigr)p_x(v_z^+) - \frac{i}{4}\Bigl(d_+(e_z^-) - d_-(e_z^-)\Bigr)p_x(v_z^-) \\
	\Delta_zE_z\bigl(S^x, v\bigr) &= 0
\end{align}
while the components with $b=y$ are given by
\begin{align}
	\Delta_zE_x\bigl(S^y, v\bigr) &= \frac{i}{4}\Bigl(d_+(e_z^+) - d_-(e_z^-)\Bigr)p_y(v_z^+) + \frac{i}{4}\Bigl(d_+(e_z^-) - d_-(e_z^-)\Bigr)p_y(v_z^-) \\
	\Delta_zE_y\bigl(S^y, v\bigr) &= \frac{1}{4}\Bigl(d_+(e_z^+) + d_-(e_z^-)\Bigr)p_y(v_z^+) - \frac{1}{4}\Bigl(d_+(e_z^-) + d_-(e_z^-)\Bigr)p_y(v_z^-) \\
	\Delta_zE_z\bigl(S^y, v\bigr) &= 0
\end{align}
and when $b=z$, we have
\begin{align}
	\Delta_zE_x\bigl(S^z, v\bigr) &= 0 \\
	\Delta_zE_y\bigl(S^z, v\bigr) &= 0 \\
	\Delta_zE_z\bigl(S^z, v\bigr) &= \frac{p_z(v_z^+) - p_z(v_z^-)}{2} \label{DE-zzz}
\end{align}

At this point it is actually not necessary to perform any further calculations to find the remaining components of the operator \eqref{DaEb-2}, provided that we stick with the choice introduced in appendix \ref{sec:A}, where the eigenbases of $J_x$, $J_y$ and $J_z$ are related to each other by rotations corresponding to cyclic permutations of the coordinate axes. Under this choice of bases, the components of $\Delta_aE_i\bigl(S^b, v\bigr)$ with $a=x$ or $a=y$ can be deduced from \Eqs{DE-zxx}--\eqref{DE-zzz} simply by making cyclic permutations of the labels $x$, $y$ and $z$.

\subsection{Second derivatives: $\Delta_{ab}E_i\bigl(S^c, v\bigr)$}
\label{sec:second}

Let us then move on to consider the operator
\begin{equation}
	\Delta_{ab}E_i\bigl(S^c, v\bigr).
	\label{D^2E}
\end{equation}
The components with $a=b$ are defined by
\begin{equation}
	\Delta_{aa}E_i\bigl(S^b, v\bigr) = \Et_i\bigl(S^b(v_a^+), v\bigr) - 2\Et_i\bigl(S^b(v), v\bigr) + \Et_i\bigl(S^b(v_a^-), v\bigr).
	\label{DaaEb-2}
\end{equation}
Here the parallel transported flux operators associated to the two outer nodes have already been encountered in the operator \eqref{DaEb-2} in the previous section, while the operator $\Et_i\bigl(S^b(v), v\bigr)$ at the central node acts like the standard flux operator. The form of the operator \eqref{DaaEb-2} as an operator on the reduced Hilbert space can therefore be read off directly from the results found in the previous section. We have
\begin{align}
	\Delta_{zz}E_x\bigl(S^x, v\bigr) &= \frac{1}{2}\Bigl(d_+(e_z^+) + d_-(e_z^+)\Bigr)p_x(v_z^+) - 2p_x(v) + \frac{1}{2}\Bigl(d_+(e_z^-) + d_-(e_z^-)\Bigr)p_x(v_z^-) \label{DDE-zzxx} \\
	\Delta_{zz}E_y\bigl(S^x, v\bigr) &= -\frac{i}{2}\Bigl(d_+(e_z^+) - d_-(e_z^+)\Bigr)p_x(v_z^+) + \frac{i}{2}\Bigl(d_+(e_z^-) - d_-(e_z^-)\Bigr)p_x(v_z^-) \\
	\Delta_{zz}E_z\bigl(S^x, v\bigr) &= 0
\end{align}
and
\begin{align}
	\Delta_{zz}E_x\bigl(S^y, v\bigr) &= \frac{i}{2}\Bigl(d_+(e_z^+) - d_-(e_z^-)\Bigr)p_y(v_z^+) - \frac{i}{2}\Bigl(d_+(e_z^-) - d_-(e_z^-)\Bigr)p_y(v_z^-) \\
	\Delta_{zz}E_y\bigl(S^y, v\bigr) &= \frac{1}{2}\Bigl(d_+(e_z^+) + d_-(e_z^-)\Bigr)p_y(v_z^+) - 2p_y(v) + \frac{1}{2}\Bigl(d_+(e_z^-) + d_-(e_z^-)\Bigr)p_y(v_z^-) \\
	\Delta_{zz}E_z\bigl(S^y, v\bigr) &= 0
\end{align}
as well as
\begin{align}
	\Delta_{zz}E_x\bigl(S^z, v\bigr) &= 0 \\
	\Delta_{zz}E_y\bigl(S^z, v\bigr) &= 0 \\
	\Delta_{zz}E_z\bigl(S^z, v\bigr) &= p_z(v_z^+) - 2p_z(v) + p_z(v_z^-) \label{DDE-zzzz}
\end{align}
As before, the remaining components of the operator \eqref{DaaEb-2} can be obtained from the above equations by cyclic permutations of the labels $x$, $y$ and $z$.

In the case $a\neq b$, the operator \eqref{D^2E} is defined by the expression
\begin{align}
	\Delta_{ab}E_i\bigl(S^c, v\bigr) = \frac{1}{4}\biggl(\Et_i\bigl(&S^c(v_{ab}^{++}), v\bigr)_{\rm sym.} - \Et_i\bigl(S^c(v_{ab}^{+-}), v\bigr)_{\rm sym.} \notag \\
	&- \Et_i\bigl(S^c(v_{ab}^{-+}), v\bigr)_{\rm sym.} + \Et_i\bigl(S^c(v_{ab}^{--}), v\bigr)_{\rm sym.}\biggr)
	\label{DabEc-2}
\end{align}
which refers to the four nodes diagonally adjacent to the central node $v$ in the $(x^a, x^b)$ -coordinate plane (see \Fig{labels}). The symmetrized flux variable
\begin{equation}
	\Et_i\bigl(S^c(v_{ab}^{\sigma}), v\bigr)_{\rm sym.} = \frac{1}{2}\Bigl(\Et_i\bigl(S^c(v_{ab}^{\sigma}), v\bigr)_{v_{ab}^{\sigma} \to v_a^{\sigma^1} \to v} + \Et_i\bigl(S^c(v_{ab}^{\sigma}), v\bigr)_{v_{ab}^{\sigma} \to v_b^{\sigma^2} \to v}\Bigr)
	\label{Et-sym}
\end{equation}
represents an average over the two natural ways in which the parallel transport from $v_{ab}^\sigma$ can be taken, and the symbol $\sigma = ++$, $+-$, $-+$ or $--$ labels the four nodes involved in \Eq{DabEc-2}. The calculation of the action of the operator proceeds in the same way as in the previous section. When applied to a reduced spin network state, each parallel transported flux operator in \Eq{Et-sym} produces a contribution of the form
\begin{equation}
	D^{(1)}_{ai}\bigl(h_{v_{ab}^\sigma\to v' \to v}^{-1}\bigr)p_a(v_{ab}^\sigma).
	\label{}
\end{equation}
The holonomy $h_{v_{ab}^\sigma\to v' \to v}$ is now a product of two factors, associated to two different coordinate directions, and each of them has to be expressed in the basis appropriate to the direction of the corresponding edge.  Due to the resulting expressions for the components of the operator \eqref{DabEc-2} being somewhat lengthy, the details of the calculation as well as its results are presented in appendix \ref{sec:B}.

\subsection{Derivatives of the inverse flux operator}
\label{sec:inverse}

It then remains to examine the operator
\begin{equation}
	\Delta_a\EE_b^i(v) = \frac{\threedots\EEt_b^i\bigl(S(v_a^+), v\bigr)\threedots - \threedots\EEt_b^i\bigl(S(v_a^-), v\bigr)\threedots}{2}.
	\label{DaEEb-2}
\end{equation}
Here
\begin{equation}
	\EEt_b^i\bigl(S(v_a^\pm), v\bigr) = \frac{1}{2}\epsilon_{bcd}\epsilon^{ijk}\Et_j\bigl(S^c(v_a^\pm), v\bigr)\Et_k\bigl(S^d(v_a^\pm), v\bigr){\cal W}_{v_a^\pm}^{-1}
	\label{}
\end{equation}
and we use the notation with dots surrounding the operator to denote a specific factor ordering, in which the holonomy operator arising from the action of the parallel transported flux operator on the right is ordered to the left of the leftmost parallel transported flux. Letting the operator ordered in this way act on a reduced spin network state, and truncating the contributions from the flux operators at leading order, the operator takes the form
\begin{equation}
	\threedots\EEt_b^i\bigl(S(v_a^\pm), v\bigr)\threedots = \frac{1}{2}\epsilon_{bcd}\epsilon^{ijk}D^{(1)}_{cj}(h_{e_a^\pm})D^{(1)}_{dk}(h_{e_a^\pm})p_c(v_a^\pm)p_d(v_a^\pm)\frac{1}{w(v_a^\pm)}.
	\label{:EEt:}
\end{equation}
To complete the extraction of the leading terms in the action of the operator, we must, as before, transform the holonomies in \Eq{:EEt:} to the basis diagonalizing the angular momentum component $J_a$, and discard the off-diagonal matrix elements after the transformation.

As an example, we display the calculation for the case $a = b = i = z$. Expanding the sums over the contracted indices in \Eq{:EEt:}, we obtain
\begin{align} 
	\threedots\EEt_z^z\bigl(S(v_z^+), v\bigr)\threedots = \Bigl(D^{(1)}_{xx}(h_{e_z^+})D^{(1)}_{yy}(h_{e_z^+}) - D^{(1)}_{xy}(h_{e_z^+})D^{(1)}_{yx}(h_{e_z^+})\Bigr)p_x(v_z^+)p_y(v_z^+)\frac{1}{w(v_z^+)}.
	\label{EEt-zzz}
\end{align}
Here the combination $p_x(v_z^+)p_y(v_z^+)/w(v_z^+)$ reduces to $1/p_z(v_z^+)$. The matrix elements of the holonomies must then be expressed in the eigenbasis of $J_z$, but this calculation has already been performed in section \ref{sec:first}. Using \Eqs{D_xx}, \eqref{D_xy}, \eqref{D_yx} and \eqref{D_yy}, we see that
\begin{align}
	D_{xx}^{(1)}(h_{e_z^+})D_{yy}^{(1)}(h_{e_z^+}) - D_{xy}^{(1)}(h_{e_z^+})D_{yx}^{(1)}(h_{e_z^+})
	= d_+(e_z^+)d_-(e_z^+) = \Id(e_z^+).
	\label{}
\end{align}
Therefore the expression \eqref{EEt-zzz} becomes simply
\begin{equation}
	\threedots\EEt_z^z\bigl(S(v_z^+), v\bigr)\threedots = \frac{1}{p_z(v_z^+)}.
	\label{}
\end{equation}
An identical calculation shows that the operator $\threedots\EEt_z^z\bigl(S(v_z^-), v\bigr)\threedots$ becomes $1/p_z(v_z^-)$. Hence we arrive at the result
\begin{equation}
	\Delta_z\EE_z^z(v) = \frac{1}{2}\biggl(\frac{1}{p_z(v_z^+)} - \frac{1}{p_z(v_z^-)}\biggr).
	\label{}
\end{equation}
The complete results for the components of the operator \eqref{DaEEb-2} are given by
\begin{align}
	\Delta_z\EE_x^x(v) &= \frac{1}{4}\Bigl(d_+(e_z^+) + d_-(e_z^+)\Bigr)\frac{1}{p_x(v_z^+)} - \frac{1}{4}\Bigl(d_+(e_z^-) + d_-(e_z^-)\Bigr)\frac{1}{p_x(v_z^-)} \label{DEE-zxx} \\
	\Delta_z\EE_x^y(v) &= -\frac{i}{4}\Bigl(d_+(e_z^+) - d_-(e_z^+)\Bigr)\frac{1}{p_x(v_z^+)} - \frac{i}{4}\Bigl(d_+(e_z^-) - d_-(e_z^-)\Bigr)\frac{1}{p_x(v_z^-)} \\
	\Delta_z\EE_x^z(v) &= 0
\end{align}
and
\begin{align}
	\Delta_z\EE_y^x(v) &= \frac{i}{4}\Bigl(d_+(e_z^+) - d_-(e_z^+)\Bigr) \frac{1}{p_y(v_z^+)} + \frac{i}{4}\Bigl(d_+(e_z^-) - d_-(e_z^-)\Bigr) \frac{1}{p_y(v_z^-)} \\
	\Delta_z\EE_y^y(v) &= \frac{1}{4}\Bigl(d_+(e_z^+) + d_-(e_z^+)\Bigr)\frac{1}{p_y(v_z^+)} - \frac{1}{4}\Bigl(d_+(e_z^-) + d_-(e_z^-)\Bigr)\frac{1}{p_y(v_z^-)} \\
	\Delta_z\EE_y^z(v) &= 0
\end{align}
and
\begin{align}
	\Delta_z\EE_z^x(v) &= 0 \\
	\Delta_z\EE_z^y(v) &= 0 \\
	\Delta_z\EE_z^z(v) &= \frac{1}{2}\biggl(\frac{1}{p_z(v_z^+)} - \frac{1}{p_z(v_z^-)}\biggr) \label{DEE-zzz}
\end{align}
together with the equations obtained from these via cyclic permutations of $x$, $y$ and $z$.

This completes the derivation of the curvature operator on the reduced Hilbert space. The operator is defined by \Eqs{R^} and \eqref{Rv-reduced}, as well as \Eqs{DE-zxx}--\eqref{DE-zzz}, \eqref{DDE-zzxx}--\eqref{DDE-zzzz}, \eqref{DEE-zxx}--\eqref{DEE-zzz} and \eqref{DDE-xyxx}--\eqref{DDE-xyzz}, which give the explicit form of the operators $\Delta_aE_i\bigl(S^b, v\bigr)$, $\Delta_{ab}E_i\bigl(S^c, v\bigr)$ and $\Delta_a\EE_b^i(v)$ appearing in \Eq{Rv-reduced}.

\section{Expectation values in reduced basis states}
\label{sec:example}

As a concrete example of the action of the curvature operator on the reduced Hilbert space, we computed the action of the operator on the reduced spin network states, which form a basis of the reduced Hilbert space. In particular, this enables us to study expectation values of the curvature operator with respect to the reduced basis states. The calculations were performed using the symbolic computer algebra library SymPy.

When the operator ${\cal R}_v$ defined by \Eq{Rv-reduced} acts on a reduced spin network state carrying fixed spins on all of its edges, the result is a linear combination of the original state together with 180 new states in which some of the spins have been changed. While the explicit form of this state is computable, the resulting expression is rather lengthy and does not seem particularly instructive. To describe the structure of the result in a qualitative manner, let us introduce the terminology of ''central edge'' to denote the six edges that are connected to the central node $v$, and ''outer edge'' to denote the edges which connect the other endpoint of an inner edge to one of the nodes $v_{ab}^\sigma$ featured in the definition of the second derivative operator $\Delta_{ab}E_i\bigl(S^c, v\bigr)$. The 180 basis states entering the result in addition to the original state can then be classified into four categories as follows:
\begin{itemize}
	\item 12 states in which the spin on one central edge has been raised or lowered by one unit;
	\item 12 states in which the spin on one central edge has been raised or lowered by two units;
	\item 60 states in which the spins on two central edges have been raised or lowered by one unit (independently of each other);
	\item 96 states in which the spins on one central edge and one outer edge have been raised or lowered by one unit independently of each other, and in such a way that the two edges whose spins have changed form a path from the central node $v$ to one of the nodes $v_{ab}^\sigma$.
\end{itemize}

When computing expectation values of the curvature operator in reduced spin network states, we work with the non-symmetric operator
\begin{equation}
	R_v = {\cal R}_v{\cal V}_v^{-1}.
	\label{RV^-1}
\end{equation}
In general, the real part of the expectation value will then correspond to the expectation value of the symmetrized operator $\tfrac{1}{2}\bigl(R_v + R_v^\dagger\bigr)$. However, in all the specific examples considered below, the expectation value of the non-symmetric operator \eqref{RV^-1} already turns out to be real.

\begin{figure}[t]
	\centering
	\includegraphics[scale=0.16]{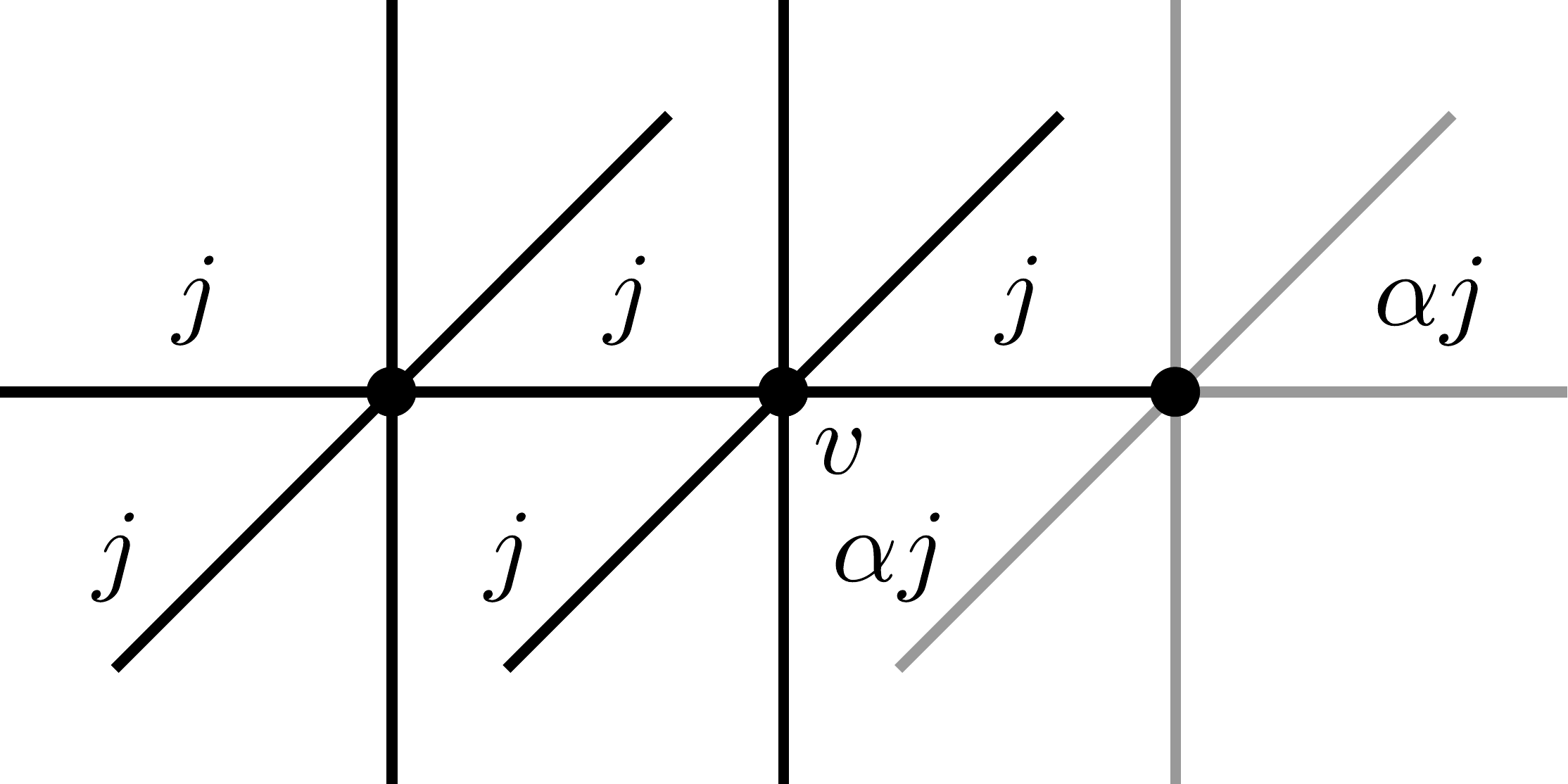}
	\caption{The reduced spin network state $\ket{\Psi_\alpha}$. The spin quantum number is $j$ on every edge, except for the five edges colored gray; these edges carry the spin $\alpha j$, where $\alpha$ is a positive constant factor.}
	\label{fig:state_alpha}
\end{figure}

\begin{figure}[t]
	\centering
	\includegraphics[scale=0.5]{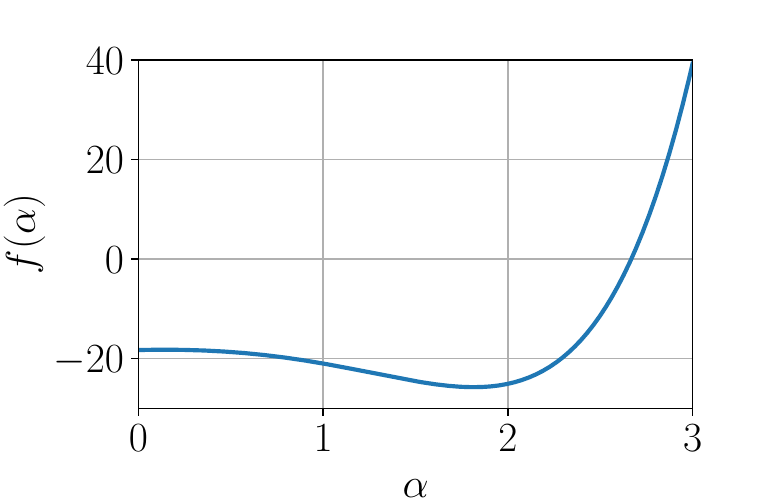}
	\caption{The function $f(\alpha)$, which characterizes the expectation value of curvature in the state $\ket{\Psi_\alpha}$.}
	\label{fig:f(a)}
\end{figure}

The simplest example of a reduced spin network state is the state $\ket{\Psi_j}$, in which every edge of the cubical graph is labeled with the same spin $j$. The expectation value of the operator $R_v$ in this state is given by
\begin{equation}
	\langle R_v\rangle_{\Psi_j} = -21\sqrt j
	\label{ev-j}
\end{equation}
(up to terms of order $1/\sqrt j$). More complicated assignments of spins can also be considered -- for example the state $\ket{\Psi_{j_xj_yj_z}}$, where every edge oriented in the coordinate direction $x^a$ carries the spin $j_a$, but the spins $j_x$, $j_y$ and $j_z$ may be different from each other. For such a state we find
\begin{equation}
	\langle R_v\rangle_{\Psi_{j_xj_yj_z}} = \frac{-7\bigl(j_x^2 + j_y^2 + j_z^2\bigr)}{\sqrt{j_xj_yj_z}}.
	\label{ev-xyz}
\end{equation}
Note that the expectation values \eqref{ev-j} and \eqref{ev-xyz} are both negative. An example of a state in which the operator $R_v$ has a positive expectation value is given by the state $\ket{\Psi_\alpha}$, which is illustrated in \Fig{fig:state_alpha}. Every edge of this state carries the spin $j$, except for the five edges $e_y^+(v_x^+)$, $e_y^-(v_x^+)$, $e_z^+(v_x^+)$, $e_z^-(v_x^+)$ and $e_x^+(v_x^+)$ incident to the node $v_x^+$. To these five edges there is assigned the spin $\alpha j$, where $\alpha$ is a positive constant factor. In the state $\ket{\Psi_\alpha}$, the expectation value of curvature is
\begin{equation}
	\langle R_v\rangle_{\Psi_\alpha} = f(\alpha)\sqrt j,
   \label{ev-a}
\end{equation}
where
\begin{equation}
	f(\alpha) = \frac{3\alpha^7 + 9\alpha^6 + 5\alpha^5 - 45\alpha^4 - 103\alpha^3 - 71\alpha^2 - 557\alpha - 585}{32(\alpha+1)}.
	\label{f(a)}
\end{equation}
A plot of the function $f(\alpha)$ is shown in \Fig{fig:f(a)}. The function has a zero at a point $\alpha = \alpha_0 \simeq 2.67$, and the values of the function are negative\footnote{In particular, $f(1) = -21$, reproducing the earlier result \eqref{ev-j}.} in the interval $0 < \alpha < \alpha_0$, while for $\alpha > \alpha_0$ the function takes positive values. Consequently, the expectation value of the operator $R_v$ in the state $\ket{\Psi_\alpha}$ is positive when $\alpha > \alpha_0$.

At a first sight it may seem surprising that the expectation value of curvature does not vanish in a state such as $\ket{\Psi_j}$, which one would intuitively expect to describe a homogeneous and isotropic geometry on a cubical lattice. However, one should note that in semiclassical terms, a reduced spin network state is sharply peaked on the intrinsic geometry represented by the flux operators, but is widely spread with respect to the extrinsic geometry encoded in the holonomy operators. For an operator involving both holonomy and flux operators, such as our curvature operator, there is therefore no especially compelling reason to insist that the expectation value of the operator in a state like $\ket{\Psi_j}$ should behave strictly according to intuitive expectations derived from the visual appearance of the state.

On the other hand, a simple quantitative estimate of the extent to which negative values of the curvature are favoured by the state $\ket{\Psi_j}$ can be obtained by considering the uncertainty (standard deviation)
\begin{equation}
	\Delta_\Psi R_v = \sqrt{\langle R_v^2\rangle_\Psi - \langle R_v\rangle^2_\Psi},
	\label{}
\end{equation}
which gives a rough characterization of the width of the probability distribution around the expectation value $\langle R_v\rangle_\Psi$. In the state $\ket{\Psi_j}$, we find
\begin{equation}
	\Delta_{\Psi_j} R_v = \frac{3\sqrt{94}}{4}\sqrt j \simeq 7.3\sqrt j.
	\label{}
\end{equation}
Comparing this with the magnitude of the expectation value $\langle R_v\rangle_{\Psi_j} = -21\sqrt j$ does suggest that, in spite of the generic arguments given above, the probability distribution of curvature in the state $\ket{\Psi_j}$ is likely to be largely concentrated on the negative side of the spectrum.

The most satisfactory way to further clarify this issue would be through a detailed semiclassical analysis of the curvature operator, in which one would evaluate expectation values of the operator in coherent states which are properly peaked on both intrinsic and extrinsic geometry. These calculations would reveal, for example, whether the curvature operator has a vanishing expectation value (up to corrections of order $\hbar$) with respect to a coherent state peaked on a flat classical geometry. On the basis of the simple examples presented above, it is difficult to anticipate whether such an analysis would indicate that the operator considered in this article is too strongly skewed towards negative values of curvature. Since we are not able to rule out the possibility that further work may show the answer to this question to be in the affirmative, let us briefly discuss a possible modified definition of the curvature operator, which we believe would be sufficient to resolve the potential problem.

If we consider the different terms in the definition of the curvature operator one by one, we find that by far the largest negative contribution to the expectation value \eqref{ev-j} is given by the second term in \Eq{Rv-reduced}, namely
\begin{align}
	2\frac{p_a(v)^2}{p_b(v)}\Delta_{aa}E_b\bigl(S^b, v\bigr) &= -8p_a(v)^2
	+\frac{p_a(v)^2}{p_b(v)}\Bigl(d_+(e_a^+) + d_-(e_a^+)\Bigr)p_b(v_a^+) \notag \\
	&\quad +\frac{p_a(v)^2}{p_b(v)}\Bigl(d_+(e_a^-) + d_-(e_a^-)\Bigr)p_b(v_a^-),
	\label{2nd-term}
\end{align}
where a sum over $a$ is implied also in the first term on the right-hand side. The expectation value of this term (multiplied with ${\cal V}_v^{-1}$) in the state $\ket{\Psi_j}$ is equal to $-24\sqrt j$. Clearly this expectation value arises entirely from the (negative-definite) first term on the right-hand side of \Eq{2nd-term}. The other terms contain the operators $d_+(e)$ and $d_-(e)$, which act by raising and lowering the spins on the respective edges, and which therefore give a vanishing contribution to the expectation value in a basis state carrying a fixed spin on every edge.

Recall now that the operator $\Delta_{aa}E_i\bigl(S^b, v\bigr)$ represents a quantization of the second covariant derivative $\D_a^2E^b_i(v)$. The first term on the right-hand side of \Eq{2nd-term} corresponds to the term $-2f(x)$ in the discretization
\begin{equation}
	f''(x) \simeq \frac{f(x+\epsilon) - 2f(x) + f(x-\epsilon)}{\epsilon^2}
	\label{}
\end{equation}
of a second derivative. This seemingly problematic term could therefore be eliminated by choosing an alternative discretization, which avoids using the central point $x$ (at the cost of having to use four total points instead of three). An example of such a discretization is given by
\begin{equation}
	f''(x) \simeq \frac{f(x+2\epsilon) - f(x+\epsilon) - f(x-\epsilon) + f(x-2\epsilon)}{3\epsilon^2}.
	\label{}
\end{equation}
The corresponding modified definition of the operator $\Delta_{aa}E_i\bigl(S^b, v\bigr)$ would be
\begin{align}
	\Delta_{aa}E_i\bigl(S^b, v\bigr) = \frac{1}{3}\Bigl(\Et_i\bigl(S^b(&v_a^{++}), v\bigr) - \Et_i\bigl(S^b(v_a^+), v\bigr) \notag \\
	&- \Et_i\bigl(S^b(v_a^-), v\bigr) + \Et_i\bigl(S^b(v_a^{--}), v\bigr)\Bigr),
	\label{}
\end{align}
where $v_a^{++}$ and $v_a^{--}$ denote the nodes that come respectively after $v_a^+$ and before $v_a^-$ in the direction of the $x^a$ -coordinate axis. If this definition is used instead of the original definition \eqref{DaaEb} in the construction of the curvature operator, the expectation value of the operator in the state $\ket{\Psi_j}$, while not exactly vanishing, would differ from zero by less than a single standard deviation.

\section{Conclusions}

In this article we studied the scalar curvature operator introduced in our previous article \cite{part1} in the setting of quantum-reduced loop gravity. We derived the explicit form of the curvature operator for quantum-reduced loop gravity by studying the action of the operator on the basis states which span the Hilbert space of the quantum-reduced model. Keeping only the terms of leading order in the spin quantum numbers in the resulting expressions, and discarding terms of lower order in $j$ (these lower-order terms generally do not belong to the reduced Hilbert space), we obtained an expression representing the curvature operator as an operator on the Hilbert space of quantum-reduced loop gravity. This operator is built out of reduced flux operators, whose action on the reduced Hilbert space is diagonal, and reduced holonomy operators, which act by raising and lowering the spin quantum numbers of a reduced spin network state in steps of $1$, and which therefore seem somewhat analogous to the raising and lowering operators of the harmonic oscillator.

As a simple example of a calculation which can be performed with the curvature operator in the quantum-reduced model, we considered expectation values of curvature in reduced spin network states. We discovered that the expectation values of the curvature operator tend to be rather strongly skewed towards the negative side of the spectrum in certain examples where one would {\em a priori} not expect either sign of the curvature to be significantly favoured. However, considering that the reduced spin network states are strongly peaked with respect to flux operators but widely spread with respect to holonomy operators, it seems unclear whether one should expect these states to satisfy one's intuitive expectations regarding operators such as the curvature operator, which are constructed out of both holonomy and flux operators.

At this stage we are therefore unable to offer a definite answer to the question of whether our results indicate any serious problem with the curvature operator. This question should be properly resolved through a systematic semiclassical analysis of the operator, where one would study expectation values of the operator in coherent states having well-defined peakedness properties with respect to both holonomy and flux operators. In this way one could verify whether the expectation values of the curvature operator agree with the expected classical results in states peaked on a given classical configuration (\eg a flat geometry). Not being able to rule out the possibility of further analysis revealing that our operator is too strongly biased towards negative values of the curvature, we proposed a slightly modified definition of the curvature operator, which we believe could help to resolve the issue. In practical terms, the modification amounts to eliminating a certain term which gives a particularly large negative contribution to the expectation value, and replacing it with a term having a vanishing expectation value with respect to reduced spin network states.

Finally, as somewhat of a side remark, our calculations seem to offer an argument to fix the undetermined multiplicative factor $\kappa_0$ in the definition of the Ashtekar--Lewandowski volume operator \cite{volume}. Another argument that has been considered in the literature is based on a consistency check between the fundamental flux operator and a so-called alternative flux operator, whose construction makes use of the fact that the inverse triad $e_a^i$ can be expressed in terms of a Poisson bracket between the Ashtekar connection and the volume. Different versions of the calculation, leading to different results for the factor $\kappa_0$, have been performed in \cite{consistency1, consistency2} and in \cite{consistency-new}. In our case, the volume operator enters the definition of the operator $\EEt_a^i\bigl(S(v), v'\bigr)$ given by \Eq{EEt}. The value of $\kappa_0$ is then determined by the requirement that the inverse flux operator ${\cal E}_a^i(v) \equiv \EEt_a^i\bigl(S(v), v\bigr)$ must act as the inverse of the flux operator $E_i\bigl(S^a(v)\bigr)$ at leading order in $j$, when these operators are applied to states in the reduced Hilbert space. This fixes the coefficient uniquely as $\kappa_0 = 1$, which is in agreement with the result originally found by Giesel and Thiemann in \cite{consistency1, consistency2}.

\begin{center}
\large{\bf{Acknowledgments}}
\end{center}

\noindent This work was supported by grant no. 2018/30/Q/ST2/00811 of the Polish National Science Center.

\appendix

\section{$SU(2)$ and angular momentum}
\label{sec:A}

In this appendix we recall a number of elementary facts from the quantum theory of angular momentum, regarding particularly the eigenstates of the angular momentum operator in the $j=1$ representation, which play an essential role in the calculations carried out in this work.

\subsection{The angular momentum operator}

The angular momentum operator is a Hermitian vector operator whose Cartesian components satisfy the commutation relation
\begin{equation}
	[J_i, J_j] = i\epsilon\downup{ij}{k}J_k
	\label{}
\end{equation}
which encodes the geometrical interpretation of the angular momentum as a generator of rotations in three-dimensional space. The solution of the eigenvalue problem can be derived solely on the basis of the commutation relation. One can simultaneously diagonalize the squared angular momentum
\begin{equation}
	J^2 = J_x^2 + J_y^2 + J_z^2
	\label{}
\end{equation}
together with one of the components, say $J_z$. The eigenvalue equations read
\begin{align}
	J^2\ket{jm} &= j(j+1)\ket{jm} \\
	J_z\ket{jm} &= m\ket{jm} \label{Jz}
\end{align}
where $j$ is any non-negative integer or half-integer, and $m$ takes values from $-j$ to $j$ in steps of $1$.

It is useful to define the raising and lowering operators
\begin{equation}
	J_\pm = J_x\pm iJ_y
	\label{}
\end{equation}
Their action on the eigenstates is given by
\begin{equation}
	J_\pm\ket{jm} = \sqrt{j(j+1)- m(m\pm 1)}\ket{j,m\pm 1}
	\label{Jpm}
\end{equation}
\Eqs{Jz}--\eqref{Jpm} specify the action of the angular momentum operator in the basis $\ket{jm}$. In particular, in the $j=1$ subspace, which is prominently featured in the present work, the components of the angular momentum operator are represented by the matrices
\begin{equation}
	J_x = \frac{1}{\sqrt 2}\begin{pmatrix} 0&1&0 \\ 1&0&1 \\ 0&1&0 \end{pmatrix}, \qquad
	J_y = \frac{1}{\sqrt 2}\begin{pmatrix} 0&-i&0 \\ i&0&-i \\ 0&i&0 \end{pmatrix}, \qquad
	J_z = \begin{pmatrix} 1&0&0 \\ 0&0&0 \\ 0&0&-1 \end{pmatrix}.
	\label{}
\end{equation}
The operator
\begin{equation}
	g(\theta, \vec n) = e^{-i\theta\vec n\cdot\vec J}
	\label{}
\end{equation}
represents a rotation around the direction $\vec n$ by the angle $\theta$. The matrices representing these operators in the subspace corresponding to the eigenvalue $j$, namely
\begin{equation}
	D^{(j)}_{mn}(g) = \bra{jm}e^{-i\theta\vec n\cdot\vec J}\ket{jn}
	\label{}
\end{equation}
define the spin-$j$ representation of $SU(2)$. The (anti-Hermitian) generators of $SU(2)$ in the spin-$j$ representation are defined as
\begin{equation}
	(\tau_i^{(j)})_{mn} = -i\bra{jm}J_i\ket{jn}
	\label{}
\end{equation}
They are normalized according to
\begin{equation}
	{\rm Tr}\,\bigl(\tau_i^{(j)}\tau_k^{(j)}\bigr) = -\frac{1}{3}j(j+1)(2j+1)\delta_{ik}
	\label{}
\end{equation}
In particular, in the fundamental representation we have
\begin{equation}
	{\rm Tr}\,(\tau_i\tau_k) = -\frac{1}{2}\delta_{ik}.
	\label{}
\end{equation}

\subsection{Eigenstates of $J_x$ and $J_y$}
\label{sec:eigenstates}

Given the eigenstates $\ket{jm}$ diagonalizing $J^2$ and $J_z$, eigenstates of $J_x$ and $J_y$ can be constructed as follows. Let $g_i$ denote the $SU(2)$ element representing a rotation which rotates the $z$-axis into the $i$-axis (where $i = x$ or $y$). Then the states
\begin{equation}
	\ket{jm}_i = D^{(j)}(g_i)\ket{jm}
	\label{jm_i}
\end{equation}
are eigenstates of the operators $J^2$ and $J_i$ with the eigenvalues $j(j+1)$ and $m$. For later use, we introduce the notation
\begin{equation}
	D^{(j)}_{mn}(g)_i = {}_i\bra{jm}D^{(j)}(g)\ket{jn}_i
	\label{}
\end{equation}
for the matrix elements of the $SU(2)$ representation matrices with respect to the basis \eqref{jm_i}.

The rotation matrix $g_i$ is not uniquely defined by the requirement of having to rotate the $z$-axis into the $i$-axis. We fix the remaining freedom by demanding that the rotation corresponds to a cyclic permutation of the coordinate axes, \ie that the rotations $g_x$ and $g_y$ rotate the axes $(x, y, z)$ respectively into $(y, z, x)$ and $(z, x, y)$. Under this choice, the eigenstates of $J_x$ and $J_y$ for $j=1$ are given by
\begin{align}
	\ket +_x &= -\frac{i}{2}\ket + - \frac{i}{\sqrt 2}\ket 0 - \frac{i}{2}\ket - \\
	\ket 0_x &= -\frac{1}{\sqrt 2}\ket + + \frac{1}{\sqrt 2}\ket - \\
	\ket -_x &= \frac{i}{2}\ket + - \frac{i}{\sqrt 2}\ket 0 + \frac{i}{2}\ket -
	\label{}
\end{align}
and
\begin{align}
	\ket +_y &= \frac{i}{2}\ket + - \frac{1}{\sqrt 2}\ket 0 - \frac{i}{2}\ket - \\
	\ket 0_y &= \frac{i}{\sqrt 2}\ket + + \frac{i}{\sqrt 2}\ket - \\
	\ket -_y &= \frac{i}{2}\ket + + \frac{1}{\sqrt 2}\ket 0 - \frac{i}{2}\ket -
	\label{}
\end{align}
where $\ket +$, $\ket 0$ and $\ket -$ are the eigenstates of $J_z$. The inverse relations expressing the eigenstates of $J_z$ in terms of those of $J_x$ and $J_y$ read
\begin{align}
	\ket + &= \frac{i}{2}\ket +_x - \frac{1}{\sqrt 2}\ket 0_x - \frac{i}{2}\ket -_x \\
	\ket 0 &= \frac{i}{\sqrt 2}\ket +_x + \frac{i}{\sqrt 2}\ket -_x \label{0.x} \\
	\ket - &= \frac{i}{2}\ket +_x + \frac{1}{\sqrt 2}\ket 0_x - \frac{i}{2}\ket -_x 
\end{align}
and
\begin{align}
	\ket + &= -\frac{i}{2}\ket +_y - \frac{i}{\sqrt 2}\ket 0_y - \frac{i}{2}\ket -_y \\
	\ket 0 &= -\frac{1}{\sqrt 2}\ket +_y + \frac{1}{\sqrt 2}\ket -_y \\
	\ket - &= \frac{i}{2}\ket +_y - \frac{i}{\sqrt 2}\ket 0_y + \frac{i}{2}\ket -_y
	\label{}
\end{align}
The choice to construct the eigenstates of the angular momentum operator in this way is made with a view towards the calculations presented in section \ref{sec:R-reduced} and appendix \ref{sec:B}, where some work can be saved by exploiting the fact that equations derived using this set of eigenstates continue to be valid when the labels $x$, $y$ and $z$ are permuted cyclically.

\subsection{Relation between the Cartesian and spherical bases}

The so-called spherical components of a vector $\vec v\in \R^3$ (see \eg \cite{Varshalovich}) are defined by their transformation properties with respect to rotations. Under a rotation corresponding to the $SU(2)$ element $g$, the components $v^m$ ($m = +1, 0, -1$) transform as
\begin{equation}
	v_m \to D^{(j)}_{nm}(g)v_n
	\label{}
\end{equation}
In terms of the Cartesian components, the spherical components of $\vec v$ are given by
\begin{align}
	v_+ &= -\frac{1}{\sqrt 2}(v_x + iv_y) \\
	v_0 &= v_z \\
	v_- &= \frac{1}{\sqrt 2}(v_x - iv_y)
	\label{}
\end{align}
Conversely, we have
\begin{align}
	v_x &= -\frac{1}{\sqrt 2}(v_+ - v_-) \label{v_x} \\
	v_y &= \frac{i}{\sqrt 2}(v_+ + v_-) \label{v_y} \\
	v_z &= v_0 \label{v_z}
\end{align}
Hence we can write down the states
\begin{align}
	\ket x &= -\frac{1}{\sqrt 2}\bigl(\ket +_z - \ket -_z\bigr) \\
	\ket y &= \frac{i}{\sqrt 2}\bigl(\ket +_z + \ket -_z\bigr) \\
	\ket z &= \ket 0_z
	\label{}
\end{align}
From the point of view of angular momentum, these are the $m=0$ eigenstates of $J_x$, $J_y$ and $J_z$ expressed in the eigenbasis of $J_z$. Expressing the same states in the eigenbases of $J_x$ and $J_y$, we obtain
\begin{align}
	\ket x &= \ket 0_x \label{x_x} \\
	\ket y &= -\frac{1}{\sqrt 2}\bigl(\ket +_x - \ket -_x\bigr) \label{y_x} \\
	\ket z &= \frac{i}{\sqrt 2}\bigl(\ket +_x + \ket -_x\bigr) \label{z_x}
\end{align}
and
\begin{align}
	\ket x &= \frac{i}{\sqrt 2}\bigl(\ket +_y + \ket -_y\bigr) \label{x_y} \\
	\ket y &= \ket 0_y \label{y_y} \\
	\ket z &= -\frac{1}{\sqrt 2}\bigl(\ket +_y - \ket -_y\bigr) \label{z_y}
\end{align}
Note that the three sets of equations above are mapped into each other by cyclic permutations of $x$, $y$ and $z$, reflecting the way in which we have chosen to construct the eigenstates of $J_x$, $J_y$ and $J_z$ in section \ref{sec:eigenstates}.

\section{Mixed components of the operator $\Delta_{ab}E_i\bigl(S^c, v\bigr)$}
\label{sec:B}

In this appendix we will derive the form of the mixed (\ie $a\neq b$) components of the discretized derivative operator
\begin{equation}
	\Delta_{ab}E_i\bigl(S^c, v\bigr)
	\label{DabEc-3}
\end{equation}
when taken as an operator on the reduced Hilbert space. If we perform the calculation using the eigenbases of $J_x$, $J_y$ and $J_z$ introduced in appendix \ref{sec:A}, we can again take advantage of the symmetry under cyclic permutations to reduce the number of components that need to be computed explicitly. Taking into account that the operator \eqref{DabEc-3} is symmetric in $a$ and $b$ by construction, we see that it suffices to take a single fixed value of the pair of indices $ab$, say $a=x$ and $b=y$, while considering all values of $i$ and $c$. From now on we will therefore consider the operator
\begin{align}
	\Delta_{xy}E_i\bigl(S^a, v\bigr) = \frac{1}{4}\biggl(\Et_i\bigl(&S^a(v_{xy}^{++}), v\bigr)_{\rm sym.} - \Et_i\bigl(S^a(v_{xy}^{+-}), v\bigr)_{\rm sym.} \notag \\
	&- \Et_i\bigl(S^a(v_{xy}^{-+}), v\bigr)_{\rm sym.} + \Et_i\bigl(S^a(v_{xy}^{--}), v\bigr)_{\rm sym.}\biggr).
	\label{DxyEa}
\end{align}
The symmetrized flux operator is defined as
\begin{equation}
	\Et_i\bigl(S^a(v_{xy}^{\sigma}), v\bigr)_{\rm sym.} = \frac{1}{2}\Bigl(\Et_i\bigl(S^a(v_{xy}^{\sigma}), v\bigr)_{v_{xy}^{\sigma} \to v_x^{\sigma^1} \to v} + \Et_i\bigl(S^a(v_{xy}^{\sigma}), v\bigr)_{v_{xy}^{\sigma} \to v_y^{\sigma^2} \to v}\Bigr)
	\label{}
\end{equation}
where
\begin{equation}
	\sigma = \bigl(\sigma^1, \sigma^2\bigr) = (++), (+-), (-+), (--)
	\label{}
\end{equation}
and the superscripts on the parallel transported flux operators specify the route along which the parallel transport from $v_{xy}^\sigma$ to the central node $v$ is taken.

{
\setlength{\extrarowheight}{6pt}
\setlength{\tabcolsep}{12pt}

\begin{table}[t]
	\centering
	\begin{tabular}{CCC}
		\Et_i\bigl(S^a(v''), v\bigr)_{v''\to v'\to v}
		& h_{v''\to v'\to v}
		& h_{v''\to v'\to v}^{-1} \vspace{4pt} \\
		\hline
		\Et_i\bigl(S^a(v_{xy}^{++}), v\bigr)_{v_{xy}^{++} \to v_x^+ \to v}
		& h_{e_x^+}^{-1}h_{e_{yx}^{++}}^{-1}
		& h_{e_{yx}^{++}}h_{e_x^+} \\
		\Et_i\bigl(S^a(v_{xy}^{++}), v\bigr)_{v_{xy}^{++} \to v_y^+ \to v}
		& h_{e_y^+}^{-1}h_{e_{xy}^{++}}^{-1}
		& h_{e_{xy}^{++}}h_{e_y^+} \\
		\Et_i\bigl(S^a(v_{xy}^{+-}), v\bigr)_{v_{xy}^{+-} \to v_x^+ \to v}
		& h_{e_x^+}^{-1}h_{e_{yx}^{+-}}
		& h_{e_{yx}^{+-}}^{-1}h_{e_x^+} \\
		\Et_i\bigl(S^a(v_{xy}^{+-}), v\bigr)_{v_{xy}^{+-} \to v_y^- \to v}
		& h_{e_y^-}h_{e_{xy}^{+-}}^{-1} 
		& h_{e_{xy}^{+-}}h_{e_y^-}^{-1} \\
		\Et_i\bigl(S^a(v_{xy}^{-+}), v\bigr)_{v_{xy}^{-+} \to v_x^- \to v}
		& h_{e_x^-}h_{e_{yx}^{-+}}^{-1} 
		& h_{e_{yx}^{-+}}h_{e_x^-}^{-1} \\
		\Et_i\bigl(S^a(v_{xy}^{+-}), v\bigr)_{v_{xy}^{-+} \to v_y^+ \to v}
		& h_{e_y^+}^{-1}h_{e_{xy}^{-+}}
		& h_{e_{xy}^{-+}}^{-1}h_{e_y^+} \\
		\Et_i\bigl(S^a(v_{xy}^{--}), v\bigr)_{v_{xy}^{--} \to v_x^- \to v}
		& h_{e_x^-}h_{e_{yx}^{--}}
		& h_{e_{yx}^{--}}^{-1}h_{e_x^-}^{-1} \\
		\Et_i\bigl(S^a(v_{xy}^{--}), v\bigr)_{v_{xy}^{--} \to v_y^- \to v}
		& h_{e_y^-}h_{e_{xy}^{--}}
		& h_{e_{xy}^{--}}^{-1}h_{e_y^-}^{-1}
	\end{tabular}
	\caption{The eight parallel transported flux operators out of which the operator $\Delta_{xy}E_i\bigl(S^a, v\bigr)$ is constructed.}
	\label{table}
\end{table}
}

The eight parallel transported flux operators entering the definition of the operator \eqref{DxyEa} are summarized in Table \ref{table}. When an operator of the form $\Et_i\bigl(S^a(v''), v\bigr)_{v''\to v'\to v}$ is applied to a reduced spin network state, its action produces the factor
\begin{equation}
	D^{(1)}_{ai}\bigl(h_{v''\to v'\to v}^{-1}\bigr)p_a(v'').
	\label{D*p}
\end{equation}
The holonomy $h_{v''\to v'\to v}$ is a product of two holonomies, one of which is associated to an edge aligned in the $x$-direction and the other to an edge in the $y$-direction. In order to identify the terms of leading order in $j$ in the action of the operator, each holonomy must be transformed to the basis corresponding to the direction of its edge with the help of \Eqs{x_x}--\eqref{z_y}. After this has been done, the leading contribution will be given by the diagonal matrix elements of the holonomy operators.

In order to organize the calculation, it is convenient to introduce the matrix
\begin{equation}
	D^{(1)}(h_{e_i}) = \begin{pmatrix}
		D^{(1)}_{xx}(h_{e_i})_i & D^{(1)}_{xy}(h_{e_i})_i & D^{(1)}_{xz}(h_{e_i})_i \\
		D^{(1)}_{yx}(h_{e_i})_i & D^{(1)}_{yy}(h_{e_i})_i & D^{(1)}_{yz}(h_{e_i})_i \\ 
		D^{(1)}_{zx}(h_{e_i})_i & D^{(1)}_{zy}(h_{e_i})_i & D^{(1)}_{zz}(h_{e_i})_i
	\end{pmatrix},
	\label{}
\end{equation}
which represents $D^{(1)}(h_{e_i})$ in the Cartesian basis $\{\ket x, \ket y, \ket z\}$, but where the matrix elements on the right-hand side are expressed in the eigenbasis of $J_i$ (as indicated by the superscript $i$). After using \Eqs{x_x}--\eqref{z_y} to evaluate the matrix elements, we discard the off-diagonal matrix elements and identify the diagonal matrix elements $D^{(1)}_{11}(h_{e_i})_i$, $D^{(1)}_{00}(h_{e_i})_i$ and $D^{(1)}_{-1\; -1}(h_{e_i})_i$ respectively with the operators $d_+(e_i)$, $\Id(e_i)$ and $d_-(e_i)$. To express the results in a more compact form, we define the (symmetric) operators
\begin{align}
	Q(e) &= \frac{1}{2}\Bigl(d_+(e) + d_-(e)\Bigr), \\
	P(e) &= \frac{i}{2}\Bigl(d_+(e) - d_-(e)\Bigr).
	\label{}
\end{align}
We then have
\begin{equation}
	D^{(1)}(h_{e_x}) = \begin{pmatrix}
		\Id(e_x) & 0 & 0 \\
		0 & Q(e_x) & -P(e_x) \\
		0 & P(e_x) & Q(e_x)
	\end{pmatrix}
	\label{D_x}
\end{equation}
if the edge is oriented in the $x$-direction, and
\begin{equation}
	D^{(1)}(h_{e_y}) = \begin{pmatrix}
		Q(e_y) & 0 & P(e_y) \\
		0 & \Id(e_y) & 0 \\
		-P(e_y) & 0 & Q(e_y)
	\end{pmatrix}
	\label{D_y}
\end{equation}
for an edge oriented in the $y$-direction. Forming the matrix product of \eqref{D_x} and \eqref{D_y} in both possible orderings, we obtain
\begin{align}
	D^{(1)}(h_{e_x})D^{(1)}(h_{e_y}) &= \begin{pmatrix}
		Q(e_y) & 0 & P(e_y) \\
		P(e_x)P(e_y) & Q(e_x) & -P(e_x)Q(e_y) \\
		-Q(e_x)P(e_y) & P(e_x) & Q(e_x)Q(e_y)
	\end{pmatrix} \label{DxDy} \\
	\intertext{and}
	D^{(1)}(h_{e_y})D^{(1)}(h_{e_x}) &= \begin{pmatrix}
		Q(e_y) & P(e_x)P(e_y) & Q(x)P(y) \\
		0 & Q(e_x) & -P(e_x) \\
		-P(e_y) & P(e_x)Q(e_y) & Q(e_x)Q(e_y)
	\end{pmatrix}. \label{DyDx}
\end{align}
The matrix elements entering the expression \eqref{D*p} can now be read off from \Eqs{DxDy} and \eqref{DyDx}. For example, when $a=x$ the relevant matrix elements (corresponding to different values of $i$) are
\begin{align}
	D^{(1)}_{xx}(h_{e_x}h_{e_y}) &= \frac{1}{2}\Bigl(d_+(e_y) + d_-(e_y)\Bigr) \\
	D^{(1)}_{xy}(h_{e_x}h_{e_y}) &= 0 \\
	D^{(1)}_{xz}(h_{e_x}h_{e_y}) &= \frac{i}{2}\Bigl(d_+(e_y) - d_-(e_y)\Bigr)
	\label{}
\end{align}
or
\begin{align}
	D^{(1)}_{xx}(h_{e_y}h_{e_x}) &= \frac{1}{2}\Bigl(d_+(e_y) + d_-(e_y)\Bigr) \\
	D^{(1)}_{xy}(h_{e_y}h_{e_x}) &= -\frac{1}{4}\Bigl(d_+(e_x) - d_-(e_x)\Bigr)\Bigl(d_+(e_y) - d_-(e_y)\Bigr) \\
	D^{(1)}_{xz}(h_{e_y}h_{e_x}) &= \frac{i}{4}\Bigl(d_+(e_x) + d_-(e_x)\Bigr)\Bigl(d_+(e_y) - d_-(e_y)\Bigr)
	\label{}
\end{align}
depending on whether the first factor of the holonomy $h_{v''\to v'\to v}$ is oriented along the $x$-axis or the $y$-axis. When passing to the final form of the operator on the reduced Hilbert space, one has to take into account the orientation of the edges specified in Table \ref{table}, using the relations $d_+(e_i^{-1}) = d_-(e_i)$ and $d_-(e_i^{-1}) = d_+(e_i)$ whenever necessary. In this way one obtains the results given by \Eqs{DDE-xyxx}--\eqref{DDE-xyzz} below. The remaining components of the operator $\Delta_{ab}E_i\bigl(S^c, v\bigr)$ can be derived from these equations by considering cyclic permutations of $x$, $y$ and $z$, and using the fact that the operator is symmetric in $a$ and $b$.

\begin{align}
	\Delta_{xy}E_x\bigl(S^x, v\bigr) ={}&\frac{1}{16}\biggl(d_+(e_{yx}^{++}) + d_-(e_{yx}^{++}) + d_+(e_y^+) + d_-(e_y^+)\biggr)p_x(v_{xy}^{++}) \notag \\
	{}-{}&\frac{1}{16}\biggl(d_+(e_{yx}^{+-}) + d_-(e_{yx}^{+-}) + d_+(e_y^-) + d_-(e_y^-)\biggr)p_x(v_{xy}^{+-}) \notag \\
	{}-{}&\frac{1}{16}\biggl(d_+(e_{yx}^{-+}) + d_-(e_{yx}^{-+}) + d_+(e_y^+) + d_-(e_y^+)\biggr)p_x(v_{xy}^{-+}) \notag \\
	{}+{}&\frac{1}{16}\biggl(d_+(e_{yx}^{--}) + d_-(e_{yx}^{--}) + d_+(e_y^-) + d_-(e_y^-)\biggr)p_x(v_{xy}^{--}) \notag \\
	\label{DDE-xyxx}
\end{align}

\begin{align}
	\Delta_{xy}E_y\bigl(S^x, v\bigr) = &-\frac{1}{32}\Bigl(d_+(e_{yx}^{++}) - d_-(e_{yx}^{++})\Bigr) \Bigl(d_+(e_x^+) - d_-(e_x^-)\Bigr)p_x(v_{xy}^{++}) \notag \\
	&-{}\frac{1}{32}\Bigl(d_+(e_{yx}^{+-}) - d_-(e_{yx}^{+-})\Bigr) \Bigl(d_+(e_x^+) - d_-(e_x^+)\Bigr)p_x(v_{xy}^{+-}) \notag \\
	&-{}\frac{1}{32}\Bigl(d_+(e_{yx}^{-+}) - d_-(e_{yx}^{-+})\Bigr) \Bigl(d_+(e_x^-) - d_-(e_x^-)\Bigr)p_x(v_{xy}^{-+}) \notag \\
	&-{}\frac{1}{32}\Bigl(d_+(e_{yx}^{--}) - d_-(e_{yx}^{--})\Bigr) \Bigl(d_+(e_x^-) - d_-(e_x^-)\Bigr)p_x(v_{xy}^{--})
	\label{}
\end{align}

\begin{align}
	&\Delta_{xy}E_z\bigl(S^x, v\bigr) = \notag \\
	&+\frac{i}{32}\biggl[\Bigl(d_+(e_{yx}^{++}) - d_-(e_{yx}^{++})\Bigr)\Bigl(d_+(e_x^+) + d_-(e_x^+)\Bigr) + 2\Bigl(d_+(e_y^+) - d_-(e_y^+)\Bigr)\biggr]p_x(v_{xy}^{++}) \notag \\
	&+\frac{i}{32}\biggl[\Bigl(d_+(e_{yx}^{+-}) - d_-(e_{yx}^{+-})\Bigr)\Bigl(d_+(e_x^+) + d_-(e_x^+)\Bigr) + 2\Bigl(d_+(e_y^-) - d_-(e_y^-)\Bigr)\biggr]p_x(v_{xy}^{+-}) \notag \\
	&-\frac{i}{32}\biggl[\Bigl(d_+(e_{yx}^{-+}) - d_-(e_{yx}^{-+})\Bigr)\Bigl(d_+(e_x^-) + d_-(e_x^-)\Bigr) + 2\Bigl(d_+(e_y^+) - d_-(e_y^+)\Bigr)\biggr]p_x(v_{xy}^{-+}) \notag \\
	&-\frac{i}{32}\biggl[\Bigl(d_+(e_{yx}^{--}) - d_-(e_{yx}^{--})\Bigr)\Bigl(d_+(e_x^-) + d_-(e_x^-)\Bigr) + 2\Bigl(d_+(e_y^-) - d_-(e_y^-)\Bigr)\biggr]p_x(v_{xy}^{--}) \notag \\
	\label{}
\end{align}

\begin{align}
	\Delta_{xy}E_x\bigl(S^y, v\bigr) = &-\frac{1}{32}\Bigl(d_+(e_{xy}^{++}) - d_-(e_{xy}^{++})\Bigr) \Bigl(d_+(e_y^+) - d_-(e_y^+)\Bigr)p_y(v_{xy}^{++}) \notag \\
	&-{}\frac{1}{32}\Bigl(d_+(e_{xy}^{+-}) - d_-(e_{xy}^{+-})\Bigr) \Bigl(d_+(e_y^-) - d_-(e_y^-)\Bigr)p_y(v_{xy}^{+-}) \notag \\
	&-{}\frac{1}{32}\Bigl(d_+(e_{xy}^{-+}) - d_-(e_{xy}^{-+})\Bigr) \Bigl(d_+(e_y^+) - d_-(e_y^+)\Bigr)p_y(v_{xy}^{-+}) \notag \\
	&-{}\frac{1}{32}\Bigl(d_+(e_{xy}^{--}) - d_-(e_{xy}^{--})\Bigr) \Bigl(d_+(e_y^-) - d_-(e_y^-)\Bigr)p_y(v_{xy}^{--})
	\label{}
\end{align}

\begin{align}
	\Delta_{xy}E_y\bigl(S^y, v\bigr) ={} &\frac{1}{16}\biggl(d_+(e_{xy}^{++}) + d_-(e_{xy}^{++}) + d_+(e_x^+) + d_-(e_x^+)\biggr)p_y(v_{xy}^{++}) \notag \\
	{}-{}&\frac{1}{16}\biggl(d_+(e_{xy}^{+-}) + d_-(e_{xy}^{+-}) + d_+(e_x^+) + d_-(e_x^+)\biggr)p_y(v_{xy}^{+-}) \notag \\
	{}-{}&\frac{1}{16}\biggl(d_+(e_{xy}^{-+}) + d_-(e_{xy}^{-+}) + d_+(e_x^-) + d_-(e_x^-)\biggr)p_y(v_{xy}^{-+}) \notag \\
	{}+{}&\frac{1}{16}\biggl(d_+(e_{xy}^{--}) + d_-(e_{xy}^{--}) + d_+(e_x^-) + d_-(e_x^-)\biggr)p_y(v_{xy}^{--}) \notag \\
	\label{}
\end{align}

\begin{align}
	&\Delta_{xy}E_z\bigl(S^y, v\bigr) = \notag \\
	&-\frac{i}{32}\biggl[2\Bigl(d_+(e_x^+) - d_-(e_x^+)\Bigr) + \Bigl(d_+(e_{xy}^{++}) - d_-(e_{xy}^{++})\Bigr)\Bigl(d_+(e_y^+) + d_-(e_y^+)\Bigr)\biggr]p_y(v_{xy}^{++}) \notag \\
	&+\frac{i}{32}\biggl[2\Bigl(d_+(e_x^+) - d_-(e_x^+)\Bigr) + \Bigl(d_+(e_{xy}^{+-}) - d_-(e_{xy}^{+-})\Bigr)\Bigl(d_+(e_y^-) + d_-(e_y^-)\Bigr)\biggr]p_y(v_{xy}^{+-}) \notag \\
	&-\frac{i}{32}\biggl[2\Bigl(d_+(e_x^-) - d_-(e_x^-)\Bigr) + \Bigl(d_+(e_{xy}^{-+}) - d_-(e_{xy}^{-+})\Bigr)\Bigl(d_+(e_y^+) + d_-(e_y^+)\Bigr)\biggr]p_y(v_{xy}^{-+}) \notag \\
	&+\frac{i}{32}\biggl[2\Bigl(d_+(e_x^-) - d_-(e_x^-)\Bigr) + \Bigl(d_+(e_{xy}^{--}) - d_-(e_{xy}^{--})\Bigr)\Bigl(d_+(e_y^-) + d_-(e_y^-)\Bigr)\biggr]p_y(v_{xy}^{--}) \notag \\
	\label{}
\end{align}

\begin{align}
	&\Delta_{xy}E_x\bigl(S^z, v\bigr) = \notag \\
	&-\frac{i}{32}\biggl[2\Bigl(d_+(e_{yx}^{++}) - d_-(e_{yx}^{++})\Bigr) + \Bigl(d_+(e_{xy}^{++}) + d_-(e_{xy}^{++})\Bigr)\Bigl(d_+(e_y^+) - d_-(e_y^+)\Bigr)\biggr]p_z(v_{xy}^{++}) \notag \\
	&-\frac{i}{32}\biggl[2\Bigl(d_+(e_{yx}^{+-}) - d_-(e_{yx}^{+-})\Bigr) + \Bigl(d_+(e_{xy}^{+-}) + d_-(e_{xy}^{+-})\Bigr)\Bigl(d_+(e_y^-) - d_-(e_y^-)\Bigr)\biggr]p_z(v_{xy}^{+-}) \notag \\
	&+\frac{i}{32}\biggl[2\Bigl(d_+(e_{yx}^{-+}) - d_-(e_{yx}^{-+})\Bigr) + \Bigl(d_+(e_{xy}^{-+}) + d_-(e_{xy}^{-+})\Bigr)\Bigl(d_+(e_y^+) - d_-(e_y^+)\Bigr)\biggr]p_z(v_{xy}^{-+}) \notag \\
	&+\frac{i}{32}\biggl[2\Bigl(d_+(e_{yx}^{--}) - d_-(e_{yx}^{--})\Bigr) + \Bigl(d_+(e_{xy}^{--}) + d_-(e_{xy}^{--})\Bigr)\Bigl(d_+(e_y^-) - d_-(e_y^-)\Bigr)\biggr]p_z(v_{xy}^{--})
	\label{}
\end{align}

\begin{align}
	&\Delta_{xy}E_y\bigl(S^z, v\bigr) = \notag \\
	&+\frac{i}{32}\biggl[\Bigl(d_+(e_{yx}^{++}) + d_-(e_{yx}^{++})\Bigr)\Bigl(d_+(e_x^+) - d_-(e_x^+)\Bigr) + 2\Bigl(d_+(e_{xy}^{++}) - d_-(e_{xy}^{++})\Bigr)\biggr]p_z(v_{xy}^{++}) \notag \\
	&-\frac{i}{32}\biggl[\Bigl(d_+(e_{yx}^{+-}) + d_-(e_{yx}^{+-})\Bigr)\Bigl(d_+(e_x^+) - d_-(e_x^+)\Bigr) + 2\Bigl(d_+(e_{xy}^{+-}) - d_-(e_{xy}^{+-})\Bigr)\biggr]p_z(v_{xy}^{+-}) \notag \\
	&+\frac{i}{32}\biggl[\Bigl(d_+(e_{yx}^{-+}) + d_-(e_{yx}^{-+})\Bigr)\Bigl(d_+(e_x^-) - d_-(e_x^-)\Bigr) + 2\Bigl(d_+(e_{xy}^{-+}) - d_-(e_{xy}^{-+})\Bigr)\biggr]p_z(v_{xy}^{-+}) \notag \\
	&-\frac{i}{32}\biggl[\Bigl(d_+(e_{yx}^{--}) + d_-(e_{yx}^{--})\Bigr)\Bigl(d_+(e_x^-) - d_-(e_x^-)\Bigr) + 2\Bigl(d_+(e_{xy}^{--}) - d_-(e_{xy}^{--})\Bigr)\biggr]p_z(v_{xy}^{--})
	\label{}
\end{align}

\begin{align}
	\Delta_{xy}E_z\bigl(S^z, v\bigr) ={} &\frac{1}{32}\Bigl(d_+(e_{yx}^{++}) + d_-(e_{yx}^{++})\Bigr)\Bigl(d_+(e_x^+) + d_-(e_x^+)\Bigr)p_z(v_{xy}^{++}) \notag \\
	{}+{}&\frac{1}{32}\Bigl(d_+(e_{xy}^{++}) + d_-(e_{xy}^{++})\Bigr)\Bigl(d_+(e_y^+) + d_-(e_y^+)\Bigr)p_z(v_{xy}^{++}) \notag \\
	{}-{}&\frac{1}{32}\Bigl(d_+(e_{yx}^{+-}) + d_-(e_{yx}^{+-})\Bigr)\Bigl(d_+(e_x^+) + d_-(e_x^+)\Bigr)p_z(v_{xy}^{+-}) \notag \\
	{}-{}&\frac{1}{32}\Bigl(d_+(e_{xy}^{+-}) + d_-(e_{xy}^{+-})\Bigr)\Bigl(d_+(e_y^-) + d_-(e_y^-)\Bigr)p_z(v_{xy}^{+-}) \notag \\
	{}-{}&\frac{1}{32}\Bigl(d_+(e_{yx}^{-+}) + d_-(e_{yx}^{-+})\Bigr)\Bigl(d_+(e_x^-) + d_-(e_x^-)\Bigr)p_z(v_{xy}^{-+}) \notag \\
	{}-{}&\frac{1}{32}\Bigl(d_+(e_{xy}^{-+}) + d_-(e_{xy}^{-+})\Bigr)\Bigl(d_+(e_y^+) + d_-(e_y^+)\Bigr)p_z(v_{xy}^{-+}) \notag \\
	{}+{}&\frac{1}{32}\Bigl(d_+(e_{yx}^{--}) + d_-(e_{yx}^{--})\Bigr)\Bigl(d_+(e_x^-) + d_-(e_x^-)\Bigr)p_z(v_{xy}^{--}) \notag \\
	{}+{}&\frac{1}{32}\Bigl(d_+(e_{xy}^{--}) + d_-(e_{xy}^{--})\Bigr)\Bigl(d_+(e_y^-) + d_-(e_y^-)\Bigr)p_z(v_{xy}^{--}) \notag \\
	\label{DDE-xyzz}
\end{align}

\end{document}